\documentclass[a4paper,fleqn,usenatbib]{mnras}

\usepackage[T1]{fontenc}
\usepackage{ae,aecompl}

\usepackage{graphicx}	
\usepackage{amsmath}	
\usepackage{amssymb}	
\usepackage{pdflscape}
\usepackage{deluxetable}

\title[Angular momentum of dwarf galaxies]{Angular momentum of dwarf galaxies}

\author[Kurapati et al.]{
Sushma Kurapati$^{1}$\thanks{E-mail: sushma@ncra.tifr.res.in},
Jayaram N. Chengalur $^{1}$, Simon Pustilnik $^{2}$, and Peter Kamphuis  $^{1}$  \\
$^{1}$ National Centre for Radio Astrophysics, Tata Institute of Fundamental Research, PO Box 3, Pune 411007, India \\
$^{2}$ Special Astrophysical Observatory, Russian Academy of Sciences, Nizhnii Arkhyz, 369167 Russia
}

\date{Accepted XXX. Received YYY; in original form ZZZ}

\pubyear{2017}

\begin{document}
\label{firstpage}
\pagerange{\pageref{firstpage}--\pageref{lastpage}}
\maketitle

\newcommand{\MB}{\ensuremath{\rm M_B}}
\newcommand{\mb}{\ensuremath{\rm M_b}}
\newcommand{\mhi}{\ensuremath{\rm M_{HI}}}
\newcommand{\mg}{\ensuremath{\rm M_{gas}}}
\newcommand{\msun}{\ensuremath{\rm M_{\odot}}}
\newcommand{\fatm}{\ensuremath{\rm f_{atm}}}
\newcommand{\jb}{\ensuremath{\rm j_b}}
\newcommand{\jg}{\ensuremath{\rm j_{gas}}}
\newcommand{\kms}{\ensuremath{\rm km \ s^{-1}}}
\newcommand{\kpc}{\ensuremath{\rm kpc}}

\begin{abstract}

Mass and specific angular momentum are two fundamental physical parameters of galaxies. We present measurements of the baryonic mass and specific angular momentum of 11 void dwarf galaxies derived from neutral hydrogen (H{\sc i}) synthesis data. Rotation curves were measured using 3D and 2D tilted ring fitting routines, and the derived curves generally overlap within the error bars, except in the central regions where, as expected, the 3D routines give steeper curves. The specific angular momentum of void dwarfs is found to be high compared to an extrapolation of the trends seen for higher mass bulge-less spirals, but comparable to that of other dwarf irregular galaxies that lie outside of voids. As such, our data show no evidence for a dependence of the specific angular momentum on the large scale environment. Combining our data with the data from the literature, we find a baryonic threshold of $\sim 10^{9.1}~M_{\odot}$ for this increase in specific angular momentum. Interestingly, this threshold is very similar to the mass threshold below which the galaxy discs start to become systematically thicker. This provides qualitative support to the suggestion that the thickening of the discs, as well as the increase in specific angular momentum, are both results of a common physical mechanism, such as feedback from star formation. Quantitatively, however, the amount of star formation observed in our dwarfs appears insufficient to produce the observed increase in specific angular momentum. It is hence likely that other processes, such as cold accretion of high angular momentum gas, also play a role in increasing the specific angular momentum.

\end{abstract}




\section{Introduction}
 
Two fundamental physical parameters of galaxies that are strongly correlated with their morphology and other secondary parameters are their mass (M) and specific angular momentum (j).
 For example, \citet{fall83} showed that spiral and elliptical galaxies occupy distinct regions in the $j_{\ast}-M_{\ast}$ plane (where the subscript indicates quantities pertaining to the stellar component of the galaxies). For both spirals and ellipticals j$_{\ast}$ $\approx$ qM$_{\ast}^{2/3}$, but at a given M$_{\ast}$ the specific stellar angular momentum of spirals is approximately 5 times larger than that of ellipticals. More recently, \citet[][hereafter OG14]{obreschkow14} using observations of all the major baryonic components in a sample of spiral galaxies, found a tight correlation between \jb\,  \mb\ and bulge fraction ($\beta$). Across the entire sample, they found that \jb\ $\propto$ M$_{b}^{2/3}$ (where the subscript indicates the quantities relating to the total baryonic components), however at a fixed bulge fraction they found that \jb\ $\propto$ \mb\ .
 
Both scalings, viz. $j \propto M^{2/3}$ as well as $j \propto M$ have some theoretical underpinning. Models in which protogalaxies acquire their angular momentum via tidal torques predict that the specific angular momentum of the dark matter halos scales with the mass as $j_{H}$ = k($\lambda$) M$_{H}^{2/3}$, where $\lambda$, the spin parameter of the halo, is approximately independent of mass\citep{peebles69}.  Numerical CDM simulations also confirm this scaling and find that $\lambda$ it is nearly independent of mass but has a slight dependence on environment \cite[e.g.][]{shi15}. This prediction is for the total mass and the total specific angular momentum, whereas the observations are limited to the quantities pertaining to the baryonic components of the galaxies. If one focuses only on the baryonic components, and assumes that the bulk of the mass is contained in a gaseous disk which is just marginally stable against gravitational collapse   \citep[see e.g][]{zasov74,zasov17}, then one can show that  \jg\ $\propto$ \mg\ (where the subscript indicates quantities relating to the gas disc).
 
In the models where the galaxy's specific angular momentum is set by tidal torquing, one would expect that the angular momentum too would have an environmental dependence, as the tidal torque depends on the environment. Indeed numerical simulations show that the spin does depend on the tidal field and that on the average, haloes spin faster in stronger tidal fields \citep[e.g.][]{shi15}. Cosmological simulations also find that the halo spin directions tend to correlate with the large scale structure, with low mass halo spins tending to be oriented parallel to the parent structure, while higher mass haloes have spins that are perpendicular to the parent structure \citep[e.g.][]{aragon-calvo07}. One possible reason for this mass dependence is that late time mass accretion is most significant for massive haloes, and tends to happen along the slowest collapsing direction (i.e. perpendicular to the parent structure, see e.g.\cite{wang17}). The relationship between the spin and the morphology also appears to be mass dependent, with simulations suggesting that halo spin is the most important parameter setting the morphology of dwarf galaxies \citep{rodriguez-gomez17}.

 This work aims to study the angular momentum of dwarf irregular galaxies, and in particular to compare the baryonic specific angular momentum of dwarf galaxies residing in voids with that of dwarfs lying outside voids. Dwarf galaxies are ideal candidates to study the environmental dependence of angular momentum as they are more sensitive to their environment as compared to larger spirals. For example, single dish H{\sc i} observations have shown that dwarfs in voids are systematically more gas rich than dwarfs in average density regions \citep{pustilnik16a}. Simulations also show that the fainter dwarf galaxies in voids (M$_{r}$ > -16) are typically bluer, have higher rates of star formation and specific star formation and lower mean stellar ages as compared to the dwarf galaxies in average density environments, while luminous dwarfs do not show similar trends \citep{kreckel11a}. Further, as mentioned above, simulations also suggest that the halo spin is the most important parameter in setting the morphology of dwarf galaxies, this makes it an interesting parameter to measure. Finally, since dwarf irregular galaxies are also generally gas rich, this means that measurements of the baryonic specific angular momentum and baryonic mass of dwarfs are less sensitive to the assumed stellar mass to light ratio as compared to brighter galaxies. Apart from comparing dwarf galaxies inside voids with those outside, we also compare the observed \jb\ vs \mb\ trends for dwarf galaxies with theoretical models such as those described above.

The rest of this paper is organized as follows. In \S \ref{obs}, we describe the sample selection, observations, data analysis, derivation of rotation curves and measurements of angular momentum and mass.  In \S \ref{results},  we present the \jb\ -- \mb\ relation for void dwarf galaxies and compare it with the relation for dwarf galaxies outside voids, as well as bulgeless spiral galaxies. In \S \ref{sec:dis}, we discuss and interpret our results in the context of various theoretical models and finally we conclude the paper with a summary of the key results in \S \ref{summary}.
\section{Observations $ \& $  Data analysis}
\label{obs}
\subsection{Sample}

Our primary sample is drawn from the Lynx-Cancer void galaxy sample of \citet{pustilnik11}. The Lynx-Cancer void is a nearby void, lying at a distance of only $\sim 18$~Mpc \citep{pustilnik11}. The proximity of this void allows one to study the properties of faint dwarf galaxies populating the void, unlike more distant voids, where typically only the brighter dwarfs have been studied. The Lynx-Cancer void galaxy sample \citep{pustilnik11,pustilnik16b,pustilnik16a} consists of a total of 104 galaxies and is believed to be close to complete for absolute B-band magnitude (M$_{B}$) $< $  -14~mag \citep{pustilnik11}. Single dish H{\sc i} data for these galaxies has been presented in \citet{pustilnik16a}, and has been used to show that void dwarf galaxies are systematically more gas rich (as measured in terms of the ratio of H{\sc i} mass to blue luminosity) than comparable luminosity galaxies lying in groups or regions of moderate density.  Interferometric images are available for a sub-sample of 26 galaxies, consisting of the upper quartile of gas rich galaxies (i.e. galaxies with M$_{HI}$/L$_{B}$ >1.9) based on either older  Very Large Array (VLA) archival data or from fresh Giant Meterwave Radio Telescope (GMRT) observations. Data for individual systems that are particularly interesting have been published in \citet [][]{chengalur13,chengalur15, chengalur17}.

For the current study, we require a sample where rotation curves can be reliably derived. Therefore, we have an additional set of selection criteria for the sample. We have considered only galaxies which have well-behaved velocity fields, are well resolved across the major axis (more than $\sim $ 6 beams). We exclude the galaxies that have disturbed velocity fields or have inclinations lower than 35$^{\circ}$ (see \S \ref{rotcur} for the justification of these criteria).  Ten out of 26 dwarf galaxies remain after imposition of these criteria. We also include the galaxy KK246 \citet{kreckel11}, residing in the nearby Tully void, in our sample. The parameters for these 11 galaxies are listed in Table \ref{Table1}. The columns are as follows.
column (1): galaxy name, (2): distance to the galaxy in Mpc, (3): absolute B-band magnitude (corrected for Galactic extinction), (4): H{\sc i} mass ($\times $ 10$^{7}$ M$_{\odot}$), (5): date of observation and (6): telescope used for the H{\sc i} observations.

\begin{table}
\begin{footnotesize}

\caption{Parameters of galaxies selected for this study}
\label{Table1}
\begin{tabular}{ p{1.4cm} p{0.7cm} p{0.8cm} p{1.0cm} p{1.4cm} p{0.4cm} }
\\
\hline
 Name  		& $d$ (Mpc) & \MB & $\mhi$ (10$^{7}$M$_{\odot}$) &Obs.Date& Telscope  \\ 
\hline
 KK246	    & 6.850 & -13.69 & 9.0 & 09.07.2010 & VLA  \\ 
 DDO47	    & 8.040 & -14.78 & 52.3 & 28.09.1984 &VLA  \\
 UGC4115	& 7.730 & -14.75 & 31.9 & 08.07.2004 &VLA \\ 
 UGC3501	& 10.07 & -13.32 & 8.6 & 22.11.2014&GMRT \\ 
 J0737+4724 & 10.40 & -12.50 & 1.8 & 24.11.2011&GMRT \\ 
 J0926+3343 & 10.63 & -12.90 & 5.2 & 21.08.2015&GMRT \\
 UGC5288	& 11.41 & -15.61 & 90.2 & 20.01.1999&VLA \\
 UGC4148	& 13.55 & -15.18 & 78.4 & 01.10.2015&GMRT \\
 J0630+23   & 22.92 & -15.89 & 135.1 &12.09.2015 &GMRT \\
 J0626+24   & 23.21 & -15.64 & 63.8 & 03.05.2015& GMRT \\
  J0929+1155 & 24.29 & -14.69 & 36.6 & 23.04.2015&GMRT \\
\hline
\multicolumn{6}{l}{ Notes: All the galaxies except KK246 are from Lynx-Cancer void.}\\
\multicolumn{6}{l}{All the parameters are taken from \citet{pustilnik11} }\\  
\multicolumn{6}{l}{and from \citet{kreckel11} for the galaxies in Lynx-Cancer}\\ 
\multicolumn{6}{l}{void and KK246 respectively.}\\ 
\end{tabular}
\end{footnotesize}
\end{table}  

\subsection{Data analysis}

 Data were analyzed following the standard procedure for spectral line interferometric imaging at the GMRT (see e.g. \citet{patra16,chengalur17}). All the calibration and data reduction were carried out using standard tasks in the Astronomical Imaging Processing System (AIPS; \citet{wells85}). The data were flagged and calibrated manually for the galaxies observed with the VLA. While, for the galaxies observed with the GMRT, flagging and calibration were done using the stand-alone pipeline {\tt "flagcal"} \citep{prasad12,chengalur13a}. In some cases, some slight additional flagging was required after passing the data through the pipeline which was done manually in AIPS. All the subsequent analysis was done using various tasks in AIPS. The GMRT does not do online doppler tracking. Hence, the CVEL task was run on all the calibrated data sets in order to correct for the Doppler shift due to earth's motion. A few galaxies were observed over multiple sessions, these were combined using the task DBCON, before imaging. The calibrated visibilities were converted to images using the task IMAGR. A trial data cube was made in order to identify the line free channels. Following this, a continuum image was made using the line-free channels and the corresponding clean components subtracted from the UV data by using the task UVSUB. Any residual continuum was subtracted in the image plane using the task IMAGR.

The hybrid configuration of the GMRT \citep{swarup91} allows us to make both high and low angular resolution images from a single observation. Hence, for the galaxies observed with the GMRT, data cubes were made at various (u,v) ranges, viz.~0-5 k$\lambda$, 0-10 k$\lambda$, 0-20 k$\lambda$ and 0-30 k$\lambda$ corresponding to $\sim$ 40$^{"}$, 25$^{"}$, 15$^{"}$ and 10$^{"}$. Single resolution data cubes were made for the galaxies observed with the VLA, where the resolution of the cube depends on the array configuration of the VLA used for the observations. The data cubes were made using the task IMAGR. Moment maps were made from the data cubes using the task MOMNT. The MOMNT task can be used to create a mask to blank the images at a given cutoff level. This mask was created by convolving the original cube with a Gaussian smoothing kernel which has a dimension of roughly the clean beam size. A Hanning smoothing kernel with a width of 3 velocity channels was also applied, after which pixels below 2$\sigma$ (where the $\sigma$ corresponds to the rms in original unsmoothed data cube) were blanked out.

\subsection{Rotation curves}
\label{rotcur}

The kinematics of gas rich disc galaxies have traditionally been modelled using the ``Tilted Ring Model'' \citep{rogstad74} where the gas disc is divided into a number of annulli, each of which is assumed to be in circular motion. In principle the free parameters for each ring would be its centre, the rotation velocity and two orientation angles. In practice, for well behaved galaxies, parameters like the centre and the orientation angles may not change much with radius. This modelling yields the ``rotation curve'', i.e. the variation of the (deprojected) rotation velocity with radius. Most of the earlier modelling software available for  determining rotation curves (e.g., ROTCUR in the {\sc gipsy} package \citep[]{vanderhulst92}), works on the 2D velocity field determined from the 3D data cube (i.e with RA, DEC and velocity as axes). Velocity fields determined in this way could suffer from systematic problems, such as a flattening of the velocity gradient in the central part of the galaxy because of the finite resolution of the observations (the so called ``beam smearing'' problem). These problems become more pronounced as the number of resolution elements across the galaxy decreases and the inclination of the galaxy disc increases. The selection criteria we used (see Sec.~\ref{obs}) were tuned to reduce these problems. The inclination criteria should not have any effect on our results, but the requirement that the galaxy be sufficiently well resolved would bias us towards the more gas rich systems. We discuss this in Sec.~\ref{theory} There are now packages available (e.g., TIRIFIC, \citet{jozsa07}; 3d-BAROLO \citet{diteodoro15}) that determine the rotation curve by directly fitting to the 3-D data cube and include effects such as beam smearing. We have used both of the 2-D and 3-D approaches to derive the rotation curves for our galaxies.

We first used the FAT pipeline \citep{kamphuis15} to fit tilted ring models to the H{\sc i} data cubes for our sample galaxies. The FAT pipeline is a wrapper around TIRIFIC \citep{jozsa07} and it also uses SOFIA (The Source Finding Application is a H{\sc i} source finding pipeline used for finding and parameterizing the galaxies in H{\sc i} data cubes \citep{serra15}) to obtain  estimates of the initial parameters for the rings. The errors reported by FAT are empirical estimates determined by taking the maximum of the difference between the unsmoothed and the smoothed profile, the variation of the smoothed profiles in the Monte Carlo fitting process and a minimum default value. In case of the rotation curve, this minimum value is set as the maximum of  5 km s$^{-1}$ or $0.5\times dv/ \sqrt{ {\rm sin}(i)}$, where $dv$ is the velocity resolution and  $i$ is the inclination (see \citet{kamphuis15} for more details on error determination).

We also used the task ROTCUR from the  GIPSY software package \citep{vanderhulst92} to fit tilted ring models to the 2-D velocity fields obtained using the MOMNT task in AIPS. The initial estimates for parameters such as the inclination, position angle and systemic velocity were taken from the FAT output. We also used the DUCHAMP software package (designed to find and describe sources in 3-D spectral cubes; \citet{whiting12}) to estimate the position angle and systemic velocity, the systematic velocity was also estimated from the synthetic global profile. We found that the systemic velocities estimated from FAT, DUCHAMP and the global profile match within $\pm$3 km s$^{-1}$. The inclinations obtained from FAT are generally within $\pm$10$^{0}$ compared with optical inclinations (estimated assuming an intrinsic thickness q$_{0}$=0.4  \citep{roychowdhury13}). The reasonableness of this choice of intrinsic thickness is confirmed by the fact that the inclinations obtained via fitting of the velocity fields match those obtained from the optical morphology. The centre of the galaxy was initially estimated as the centre of symmetry of the H{\sc i} moment 0 map. None of the geometric or orientation parameters were found to have a significant variation with radius. The final rotation curve was derived after fixing all the parameters except for rotation velocity. For the GMRT data cubes, this process was repeated for the different resolution data sets, to derive rotation curves at various resolutions. Finally, a hybrid rotation curve was made by combining the different rotation curves for a given galaxy. Rotation curves were also derived at all resolutions using approaching and receding halves of the galaxy independently. Errors on the rotation curve were taken as the quadrature of the errors reported by the tilted ring model fitting routines and the difference between the rotation velocities of the approaching and receding sides of the galaxy. 

The rotation curves derived using both the FAT and ROTCUR are shown in the appendix. The rotation curves are generally in excellent agreement for 8 of
the 11 galaxies. The main difference is that, as expected, the inner parts of rotation curves derived using FAT are typically steeper than those derived using ROTCUR. For 3 of the 11 galaxies (J0737+4724, UGC 3501 and DDO47), the rotation curves derived using the FAT pipeline do not match to the rotation curves derived using the ROTCUR in the inner regions. This is probably because for all three galaxies the central  H{\sc i} density distribution is clumpy. FAT assumes a symmetric density distribution, and the clumpiness in the observed density made the central parts of the FAT rotation unreliable, as confirmed by the large errors reported by FAT. While it is possible to model asymmetric density distributions using TIRIFC, we have not attempted that here. Instead, we use the rotation curves derived using ROTCUR for these three galaxies, since, as discussed above, in general, there is not much difference between the rotation curves derived by the two packages. The systematic difference that is seen in the inner regions of the rotation curve will have negligible effect on the total angular momentum (which is dominated by the outer parts of the disc) which is the physical parameter that is of primary interest to us here.

\subsection{Measurement of angular momentum and mass}
\begin{table*}
\begin{footnotesize}

\caption{Measured values of mass and specific angular momentum of 11 galaxies in this study  }
\label{table2}
\begin{tabular}{ p{0.3cm} p{1.65cm}  p{1.7cm}  p{1.7cm} p{1.7cm} p{2.45cm}  p{2.45cm} p{2.45cm} }
\\
\hline
 ID & Galaxy  & M$_{\ast}$ & M$_{gas}$ & M$_{bar}$ & j$_{\ast}$ & j$_{gas}$ & j$_{bar}$\\ 
    & &(log$_{10}$ M$_{\odot}$) & (log$_{10}$ M$_{\odot}$) & (log$_{10}$ M$_{\odot}$) & (log$_{10}$ kpc km s$^{-1}$) & (log$_{10}$ kpc km s$^{-1}$) & (log$_{10}$ kpc km s$^{-1}$) \\
\hline
 1& KK246	   &7.70$^{+0.114} _{-0.155}$ &8.11 $^{+0.119} _{-0.165}$ &8.25 $^{+0.094} _{-0.121}$ &1.10$^{+0.081}_{-0.100} $ &1.96 $^{+0.062}_{-0.073} $& 1.84 $ ^{+0.062} _{-0.073} $\\[1.0ex]
 2& DDO47	   &7.55$^{+0.114} _{-0.155}$ &8.85 $^{+0.119} _{-0.165}$ &8.87 $^{+0.114} _{-0.156}$ &2.06$^{+0.062}_{-0.072} $ &2.43 $^{+0.061}_{-0.071} $& 2.42 $ ^{+0.061} _{-0.071} $\\[1.0ex]
 3& UGC4115	   &7.57$^{+0.114} _{-0.155}$ &8.71 $^{+0.119} _{-0.165}$ &8.74 $^{+0.112} _{-0.152}$ &1.74$^{+0.073}_{-0.088} $ &2.11 $^{+0.064}_{-0.075} $& 2.09 $ ^{+0.064} _{-0.075} $\\[1.0ex]
 4& UGC3501	   &7.09$^{+0.121} _{-0.168}$ &8.07 $^{+0.119} _{-0.165}$ &8.11 $^{+0.110} _{-0.147}$ &1.36$^{+0.062}_{-0.073} $ &1.45 $^{+0.063}_{-0.073} $& 1.44 $ ^{+0.063} _{-0.073} $\\[1.0ex]
 5& J0737+4724 &6.07$^{+0.114} _{-0.155}$ &7.38 $^{+0.119} _{-0.165}$ &7.41 $^{+0.115} _{-0.156}$ &0.92$^{+0.063}_{-0.074} $ &1.22 $^{+0.063}_{-0.074} $& 1.21 $ ^{+0.063} _{-0.074} $\\[1.0ex]
 6& J0926+3343 &6.09$^{+0.114} _{-0.155}$ &7.85 $^{+0.119} _{-0.165}$ &7.86 $^{+0.118} _{-0.162}$ &1.33$^{+0.080}_{-0.098} $ &1.52 $^{+0.073}_{-0.088} $& 1.52 $ ^{+0.073} _{-0.088} $\\[1.0ex]
 7& UGC5288	   &8.14$^{+0.114} _{-0.155}$ &9.08 $^{+0.119} _{-0.165}$ &9.13 $^{+0.109} _{-0.146}$ &2.16$^{+0.062}_{-0.072} $ &2.73 $^{+0.061}_{-0.071} $& 2.70 $ ^{+0.061} _{-0.071} $\\[1.0ex]
 8& UGC4148	   &7.46$^{+0.114} _{-0.155}$ &9.10 $^{+0.119} _{-0.165}$ &9.11 $^{+0.117} _{-0.161}$ &2.24$^{+0.062}_{-0.072} $ &2.53 $^{+0.061}_{-0.071} $& 2.53 $ ^{+0.061} _{-0.071} $\\[1.0ex]
 9& J0630+23   &8.22$^{+0.115} _{-0.157}$ &9.29 $^{+0.119} _{-0.165}$ &9.33 $^{+0.111} _{-0.150}$ &2.22$^{+0.063}_{-0.073} $ &2.67 $^{+0.061}_{-0.071} $& 2.65 $ ^{+0.061} _{-0.071} $\\[1.0ex]
 10& J0626+24  &7.73$^{+0.115} _{-0.157}$ &8.95 $^{+0.119} _{-0.165}$ &8.97 $^{+0.114} _{-0.154}$ &2.37$^{+0.062}_{-0.072} $ &2.41 $^{+0.062}_{-0.072} $& 2.41 $ ^{+0.062} _{-0.072} $\\[1.0ex]
 11&J0929+1155 &7.43$^{+0.114} _{-0.155}$ &8.68 $^{+0.119} _{-0.165}$ &8.70 $^{+0.114} _{-0.155}$ &2.23$^{+0.063}_{-0.073} $ &2.32 $^{+0.062}_{-0.073} $& 2.32 $ ^{+0.062} _{-0.073} $\\[1.0ex]
 \hline & 
\end{tabular}

\end{footnotesize}

\end{table*}

For all of our sample galaxies, the H{\sc i} mass was calculated using the standard formula , M$_{HI} $= 2.36$\times$ 10$^{5}$ $D^{2} \int$S $dv~ M_{\odot}$, where $D$ is the distance in Mpc, $S$ is the flux density in Jy and $dv$ is in km s$^{-1}$ \citep{roberts62}. Integrated fluxes for the galaxies observed at the GMRT  were measured using the lowest resolution ($\sim$ 40$^{"}$) data. The H{\sc i} flux was measured both from the integrated global profile, as well as directly from the data cube using the SOFIA software. These two are in excellent agreement (within $\sim$   10\%); the value quoted here is from SOFIA. The quoted uncertainties on gas mass are based on the errors on the distance measurements (typically 10$\%$) as well as the error on flux densities. The error on the flux density is dominated by the calibration uncertainties, which for the GMRT are typically $\sim$ 10\%.

The specific angular momentum j can be estimated by evaluating

\begin{equation}
j \equiv \frac{J}{M} = \frac{\int_{0}^{R}dr 2\pi r^{2} \Sigma(r) v(r) cos[\delta i(r)]}{\int_{0}^{R}dr 2\pi r \Sigma(r)}
\end{equation}

where, $\Sigma(r)$ is the azimuthally averaged surface density, v(r) is the circular velocity at radius r, $\delta$i(r) is the difference in inclination with respect to the central disc inclination and R is the radius of the gas disk. This integral needs to be evaluated separately for the stellar and the gaseous components of the galaxy. 
The H{\sc i}  and optical deprojected radial surface brightness profiles were derived by fitting elliptical annuli to the H{\sc i} moment~0 map and g-band optical images respectively, by using the ELLINT task in the GIPSY package. The g-band optical images were taken from the SDSS \citep{ahn12} or from PanSTARRS \citep{flewelling16} for those galaxies which lie outside the SDSS footprint.  The stellar masses for these galaxies were calculated  using the g-band luminosities and the g-i colors, which were taken from \citet{perepelitsyna14}\footnote{We note that there are errors in calculations of stellar masses in  \cite{perepelitsyna14}, where the galactic extinctions were not taken into account. We have hence recalculated the stellar masses for  this paper. An updated version of stellar masses in \cite{perepelitsyna14}  is now given in arXiv:1408.0613v3.}, and using the stellar mass to light relations given in \citet{zibetti09}. 
Errors in the stellar mass were computed using the errors in the distance as well as the photometric errors. The total gas density profile was computed by scaling the H{\sc i} surface density by a factor of 1.35 to account for the contribution from Helium. No correction was made for the molecular gas since for faint dwarf galaxies such as those in our sample, the molecular fraction is believed to be low \citep[e.g. ][]{taylor98,schruba12, cormier14}. The contribution of the ionized gas is also expected to be small and has been neglected. Thus, the total baryon density profiles were computed as $\Sigma_{b} = \Sigma_{\ast} + 1.35$  $\Sigma_{HI}$. The specific angular momentum of gas, stars and baryons was computed by numerically evaluating the integral in Equation (1). For all quantities, the integral converges within the error bars. The rotation velocity of the stars was taken to be the same as that of the gas. This assumption has no significant effect on our results since the total angular momentum of the stellar component is very small (< 10 $\%$) compared to that of the gas. Table \ref{table2} lists the measurements of total mass and specific angular momentum of stars, gas, and baryons along with error bars on these quantities. The columns are as follows. Column (1) galaxy ID;  (2) name of the galaxy; (3) stellar mass ; (4) gas mass; (5) total baryonic mass; (6) stellar specific angular momentum; (7) gas specific angular momentum and (8) baryonic specific angular momentum.


\section{Results}
\label{results}

\begin{figure}
  \centering
  \includegraphics[width=1.05\linewidth]{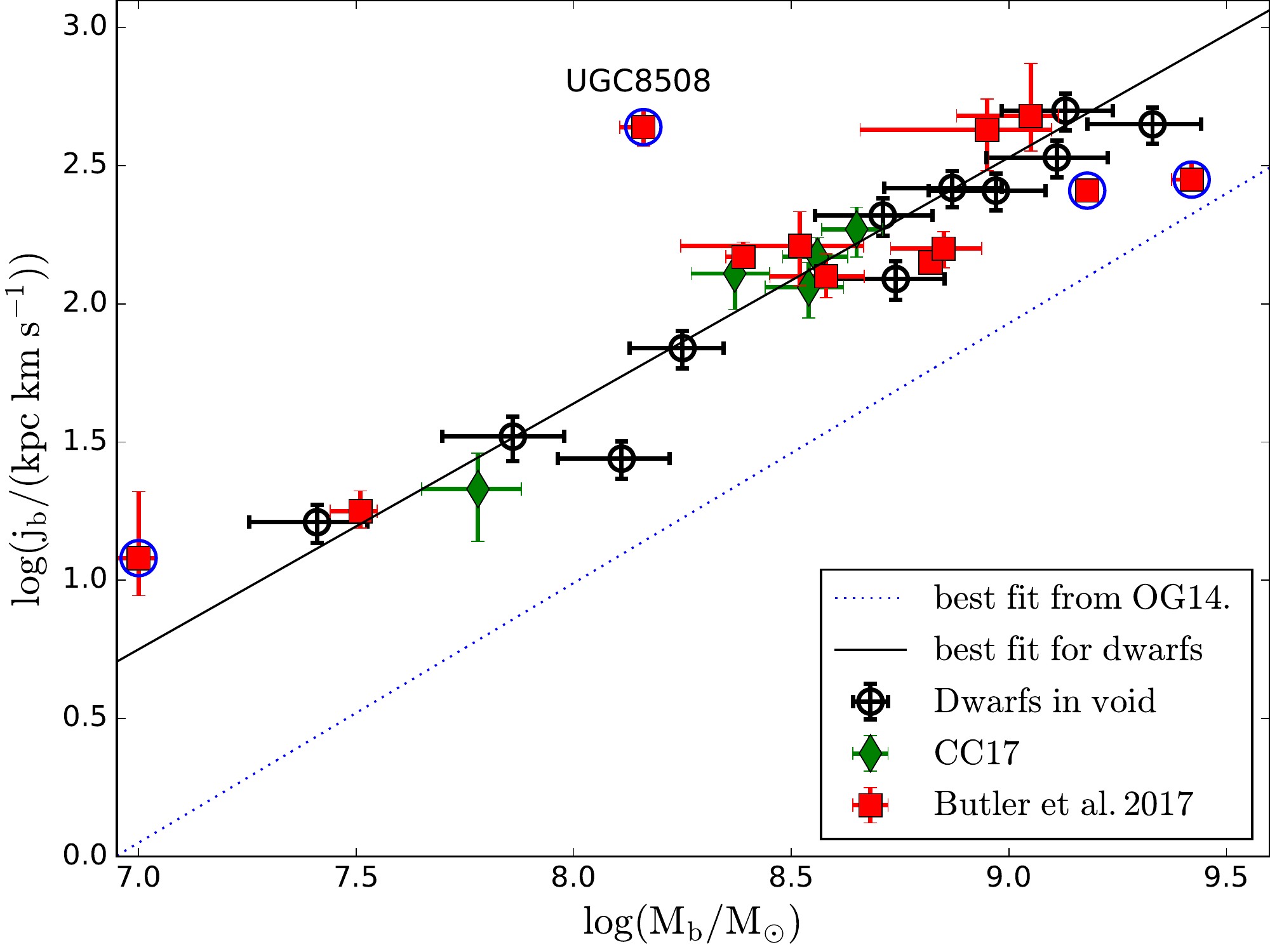}
  \caption{Log \jb -log \mb relation of 11 dwarf galaxies residing in Lynx-Cancer void from this work (black open circles), 12 dwarf galaxies from \citet{butler17} (red squares) and 5 dwarf galaxies from \citet{chowdhury17} (green diamonds). The blue dotted line indicates the $\beta$ = 0 plane of \jb - \mb relation obtained for the massive spiral galaxies by \citet{obreschkow14} and was recomputed by \citet{chowdhury17}. The black solid line is the best fit line for the dwarf galaxies using the linear regression. The galaxies from \citet{butler17} that were identified as being discrepant are marked with a circle around the red squares.}
  \label{angmom}
\end{figure}

\subsection{Comparison of j in void dwarfs $\&$ other dwarfs}
\label{j-voids}

We show in Fig.~\ref{angmom} the {\jb-\mb} relation for the 11 void dwarf galaxies (black open circles). One of the aims of this study was to check if void galaxies have a different specific angular momentum compared to galaxies outside voids. Hence, we also show in Fig.~\ref{angmom} data for a comparison sample of 14 dIrr galaxies drawn from the studies by \citet{butler17} (red squares) as well as for 5 gas rich dwarf galaxies taken from \citet[][hereafter CC17]{chowdhury17} (green diamonds). Since two galaxies (DDO 154 and DDO 133) were common in both the samples, they were excluded from the \citet{butler17} sample. This gives us a total of 17 galaxies in the comparison sample. These two samples contain galaxies with similar morphological type (mostly dwarf irregulars) and \MB\  as the galaxies in our sample. While, as noted above, our sample galaxies were selected from void environments, the large scale environment was not a specific selection criteria for the two comparison samples. However, some of the galaxies from the comparison samples lie in low density regions.  In the sample of \citet{chowdhury17}, the galaxy AndIV is an isolated galaxy and lies in a low density region \citep{karachentsev16}.  In the sample of \citet{butler17}, DDO52 belongs to Lynx-Cancer void. As can be seen from the plot, the data for the void dwarfs from our sample and other dwarfs from the literature overlap. A Kolmogorov-Smirnov (KS) test gives a probability of 0.54 for the galaxies in our sample and the literature being drawn from the same distribution.  If the two galaxies from the comparison sample that happen to lie in low density regions (viz. AndIV and DDO52) are included in the void galaxy sample, the KS test  gives a probability of $\sim$ 0.30. This large p-value indicates that there is no clear statistical evidence for the two samples (dwarfs from voids and average density regions) being drawn from different populations. Hence, when we determine the {\jb-\mb} relation below, we combine the data for all the galaxies. This gives us a total sample size of 28 (11 galaxies from our sample, 12 galaxies from \citet{butler17} and 5 galaxies from \citet{chowdhury17}). We note that one of the selection criteria for all of these samples was the availability of good quality HI rotation curves. This means that while all of the samples consist of gas rich dwarfs, they do not constitute a volume limited complete sample. We return to the issue of having selected gas rich dwarfs in Section~\ref{theory}.

\subsection{j$_{b}$ - M$_{b}$ relation for dwarf galaxies}
\label{M-j}

Another useful comparison is with the sample of spiral galaxies from the THINGS survey. \citet{obreschkow14} studied the relation between specific angular momentum (\jb), baryonic mass (\mb) and bulge fraction($\beta$) for the galaxies in this sample, and found that these three parameters correlate. The dwarf galaxies in our sample are bulge-less, and for Fig.~\ref{angmom} we would like a parametrization of the relationship between \jb\ and \mb\ for $\beta=0$. Such a parameterization has been provided by \citet{chowdhury17} (based on the data in \citet{obreschkow14}, and with errors estimated via bootstrapping re-sampling), viz.

\begin{equation}
\log_{10} (\frac{\jb}{10^{3} \ \kpc \ \kms}) = c_{1} \log_{10}( \frac{\mb}{10^{10} \ \msun}) + c_{2}
\end{equation}

with (c$_{1}$, c$_{2}$) = (0.94 $\pm$ 0.05, -0.13 $\pm$ 0.04) respectively. This is shown in Fig.~\ref{angmom} as the dotted blue line. 
As can be seen, all of the dwarf galaxies lie systematically above the relationship seen for bulgeless spiral galaxies, a fact noted earlier by \citet{butler17} and \citet{chowdhury17}. Since the dwarf galaxies both inside and outside of voids appear to follow the same trend in the {\jb-\mb} plane, we combine both the void and non-void samples in order to determine the best fit {\jb-\mb} for dwarf irregular galaxies. 

We use two fitting procedures to estimate the best fit linear regression between \jb and \mb\ for the combined sample. The LTS algorithm \citep{cappellari13} does an iterative fit allowing for errors on both axes (but assuming that they are uncorrelated). In each iteration data points identified as discrepant in the previous iteration are dropped. This algorithm identifies four galaxies from the sample of \citet{butler17}, viz. UGC8508, DDO210, DDO50 and NGC2366 as being discrepant.  As can be seen from Fig.~\ref{angmom}, UGC8508 is a significant outlier. We could not identify any underlying physical reason for this. All the discrepant galaxies are indicated by a circle around the red squares in Fig \ref{angmom}. After excluding these galaxies, the final fit values that the LTS algorithm gives are (c$_{1}$, c$_{2}$) = (0.83 $\pm$ 0.06, 0.31 $\pm$ 0.09). We then use the BCES algorithm \citep{akritas96} as implemented in the python code developed by R. Nemmen \citep[see e.g.][]{nemmen12} which accounts for errors in both {\jb} and {\mb}, as well as the fact that these errors are correlated. The galaxies identified by the LTS algorithm as discrepant have been dropped when doing this fit. The BCES algorithm fit values are (c$_{1}$, c$_{2}$) = (0.89 $\pm$ 0.05, 0.40 $\pm$ 0.08). We adopt these fit values for the rest of the paper; this best fit line is also shown by the solid black line in Fig.~\ref{angmom}. Including the three discrepant galaxies, still excluding UGC 8508, gives best fit parameter values which agree with the values above within 2$\sigma$. It is interesting that the slope (i.e. the c$_{1}$ parameter) for the dwarf galaxies matches the slope measured for larger spirals within the 1$\sigma$ error bars. However, the intercept obtained for the dwarf galaxies is significantly ($\sim 6\sigma$, using the combined error bars) different, with the dwarf galaxies having a higher specific angular momentum than that expected from the relationship for spiral galaxies.

\section{Discussion}
\label{sec:dis}

In theoretical models of galaxy formation, the angular momentum acquired in proto-galaxies is presented in terms of the ``spin parameter'' ($\lambda$ = J |E|$^{1/2}$ /GM$^{5/2}$ ), which is a combination of the total angular momentum J, the total energy E and the total mass M. This parametrisation is popular since the value of $\lambda$ (unlike $j$) is expected to be independent of mass. Consistent with early theoretical calculations, simulations of structure formation find that the spin parameter of galaxy halos exhibits a log-normal distribution with average <$\lambda$> $ \approx 0.04$. The distribution of spin parameter has been studied as a function of mass, redshift, and environment, and is found to be nearly independent of mass and redshift, but with a slight dependence on environment \cite[e.g.][]{shi15}. The fact that $\lambda$ is independent of mass also means that the specific angular momentum scales with mass as j $\propto$ M$^{\alpha}$, with $\alpha \sim 2/3$, \citep{peebles69,peebles71}.   These theoretical and numerical results are strictly speaking for the dark matter haloes. Observations can only directly probe the baryonic matter. If one assumes that the baryons and the dark matter initially have the same specific angular momentum and that the specific angular momentum is roughly conserved as the baryons collapse to form the visible galaxy, then for the baryonic material too, one would observe the same scaling between specific angular momentum and total baryonic mass, viz. $j_b \propto M_b^{2/3}$. In a pioneering study done more than three decades ago, \citet{fall83} showed that for both spiral and elliptical galaxies $j_b \propto M_b^\alpha$ with $\alpha \approx 2/3$. The constant of proportionality is larger for spiral galaxies than for ellipticals by a factor $\approx 5$, a result confirmed in \cite{romanowsky12}. These authors also show that the differences between ellipticals and spirals cannot be explained by assuming that they arise in proto-galaxies with different spin parameters, instead there must be formation and evolutionary processes that affect the final specific angular momentum of the baryons.

\begin{figure}
  \centering
  \includegraphics[width=1.05\linewidth]{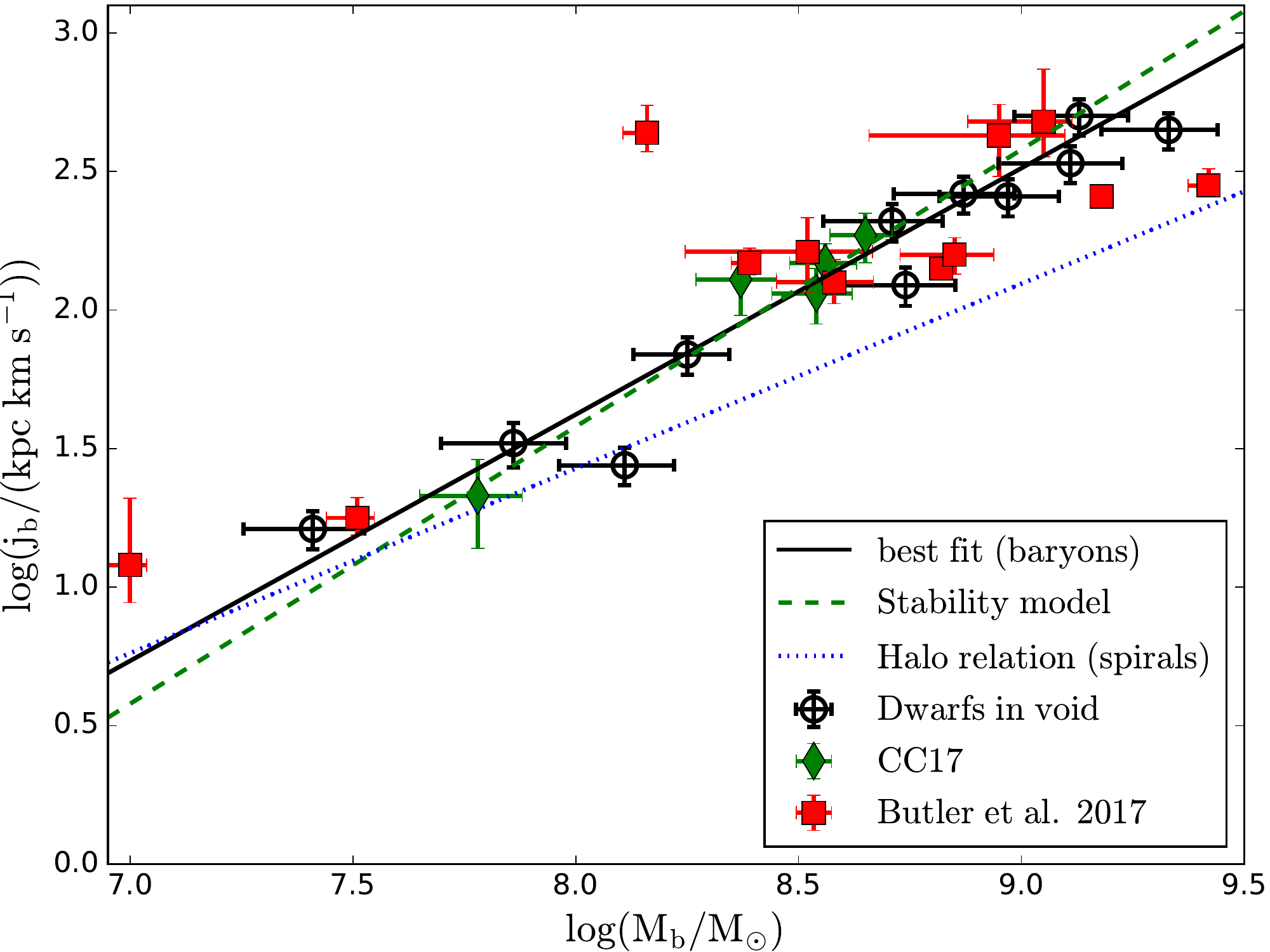}
  \includegraphics[width=1.05\linewidth]{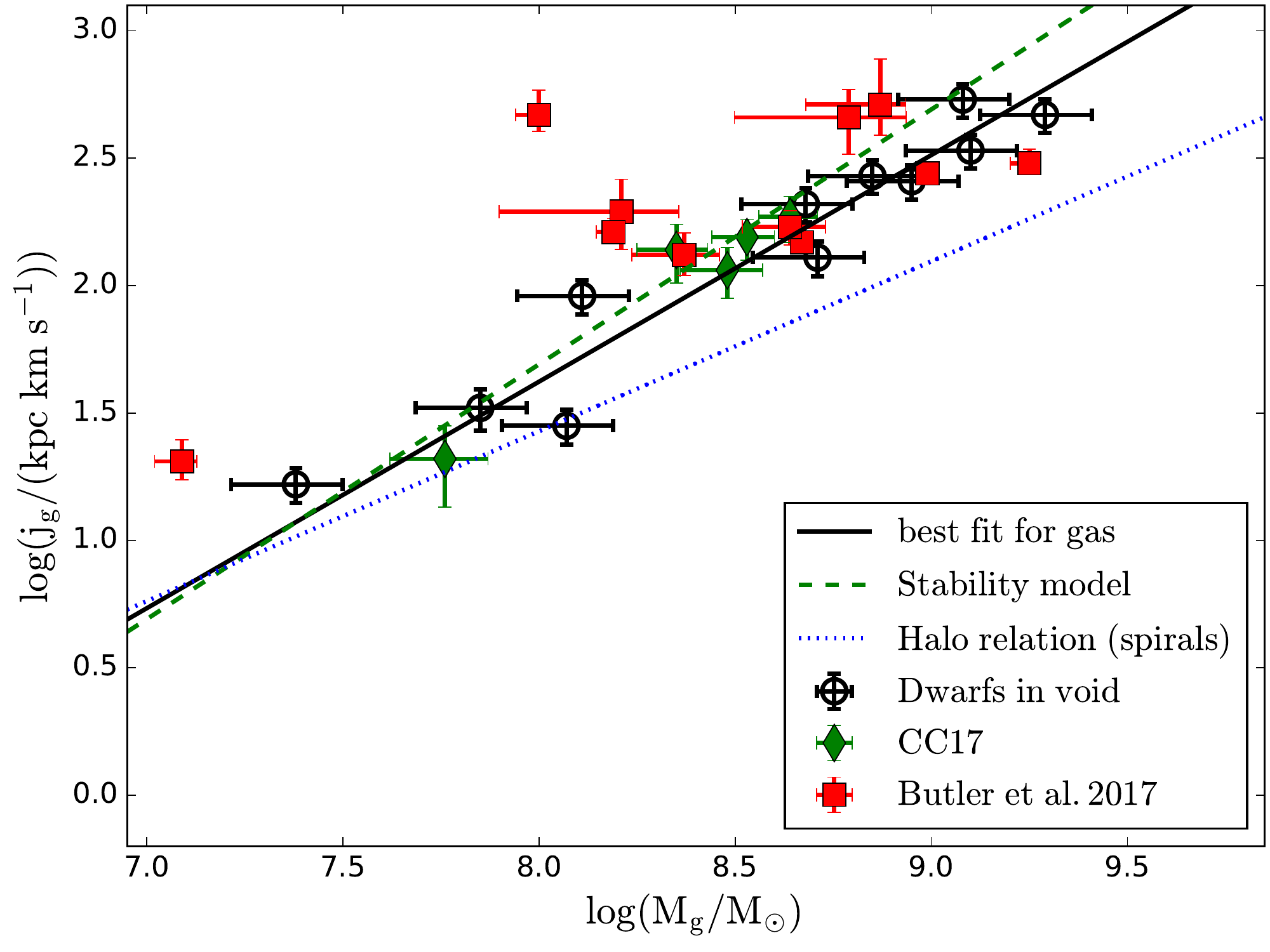}
  \caption{(a). The Log \jb - Log \mb plane and (b) Log \jg - Log \mg\  plane showing data for 11 dwarf galaxies residing in Lynx-Cancer void (black open circles) as well as dwarf galaxies for which we have taken the data from the literature. The black solid line indicates the best fit linear relation to the entire sample. The green dotted line is the best-fit stability model (Eqn~\ref{eqn:toomre}) for a rising rotation curve. The best fit stability model corresponds to $\sigma$/$Q_{c} \sim$ 2.8 \kms and $\sim$ 2.2 \kms for the fit to the baryons and the gas   respectively. The dotted line is from \citet{obreschkow14} and shows the $j \propto M^{2/3}$
 relation expected from the tidal torquing model. See the text for more details.}
  \label{angmom_stability}
\end{figure}

While halos may acquire their initial angular momentum via tidal torquing, there are a number of evolutionary processes that could affect the specific angular momentum of the galaxies we observe today. Bulge dominated objects produced by mergers of discs could have greatly reduced specific angular momentum compared to their progenitors. This is consistent with the results of \cite{fall83} who find that the specific angular momentum of ellipticals is significantly smaller than that of spirals. Feedback from star formation or AGN could prevent the catastrophic collapse of baryonic matter into a low specific angular momentum bulge in the centre of the halo. Indeed one of the major problems in the early simulations of galaxy formation that did not include such feedback was that the simulated galaxies had unrealistically small specific angular momentum \citep[e.g.][]{navarro97}. Conversely, this problem is greatly diminished in models which include feedback (see e.g. \citet{okamoto05, governato07}). If the baryonic specific angular momentum depends critically on feedback processes, rather than on the initial angular momentum of the dark halo, then it is possible that there is some kind of regulatory mechanism that results in the observed relationship between the baryonic specific angular momentum and the baryonic mass. Below we look at one specific model where it is assumed that the galaxy consists of a marginally stable gas disk. 

\subsection{Disk stability models}
\label{Stability_model}

\citet{zasov74} showed that the H{\sc i} mass (\mhi) scales with the maximum rotational velocity ($V_m$) and the radius at which it is achieved ($R_m$), i.e. $\mhi \propto V_m R_m$ (the product $V_m R_m$ can be taken as an approximate proxy to the specific angular momentum), and argued that this scaling can be understood as a consequence of a critical gas density required for the onset of star formation. This critical density is assumed to be the one set by the Toomre stability criteria \citet{toomre64}. \citet{zasov17} present analytic equations for the relationship between the specific angular momentum, mass, and the Toomre stability parameter. We summarize these briefly below. For a gas disc with radius $R_0$ and the circular velocity $V_0$ at $R_0$, and for which the column density is everywhere equal to the critical density, the gas mass is given by 

\begin{equation}
  \mg = 2^{3/2} (1+n)^{-1/2} \frac{\sigma}{Q_c G} V_0 R_0
  \label{eqn:mgcrit}
\end{equation}

where $n=0$ for a flat rotation curve, and $1$ for a rising one and $Q_c$ is the numerical value of the threshold parameter at which the disc instability sets in, in units of the Toomre parameter $Q = \sigma \kappa /(\pi G \Sigma_g)$, where $\kappa$ is the epicyclic frequency, $\sigma$ is the velocity dispersion and $\Sigma_g$ the gas column density (\citet{zasov17})\footnote{We note that there is a typographical error in \cite{zasov17}, where the factor $G$ is missing}. Similarly the
total angular momentum is given by

\begin{equation}
  J = 2^{1/2} (1+n)^{-1/2} \frac{\sigma}{Q_c G} V_0^2 R_0^2
  \label{eqn:Jcrit}
\end{equation}

This gives us the relation between the specific angular momentum and mass if the gas is in a marginally stable state, viz.
\begin{equation}
 \log(j) = \log(M) + \log \big(2^{-5/2} (1+n)^{1/2} \frac{Q_c G}{\sigma}\big)
  \label{eqn:jcrit}
\end{equation}

We note that the slope of the relation is independent of the shape of the rotation curve (flat or rising) although the intercept is not. 
Hence, if the gas is marginally stable, we expect that the specific angular momentum scales with mass as j $\propto$ M$^{\alpha}$, with $\alpha \sim 1$. This is in contrast with the expected scaling relation for the specific angular momentum and mass of the dark matter halo ($\alpha \approx 2/3$; \citep{peebles69, peebles71}) and also the observed scaling relation between the baryonic specific angular momentum and mass of the spiral galaxies and elliptical galaxies ($\alpha \approx 2/3$; \citet{fall83}), but is similar to the scaling relation obtained for galaxies at a fixed bulge fraction ($\alpha \sim 0.94 \pm 0.05$; \citet{obreschkow14}) and also for the dwarf galaxies ($\alpha \sim 0.89  \pm 0.05$; see \S \ref{M-j}). 

For the case of a linearly rising rotation curve, we can write Eqn.~\ref{eqn:jcrit}  as
\begin{equation}
 \log(j) = \log(M) + \log \bigg(\frac{Q_c G}{4\sigma}\bigg)
  \label{eqn:toomre}
\end{equation}

Since dwarf galaxies are gas dominated, Eqn.~\ref{eqn:toomre} should approximately hold for the total baryonic content. Fig \ref{angmom_stability} shows the location of 11 dwarf galaxies residing in Lynx-Cancer void (from this work; black open circles) and dwarf galaxies from the literature in the (a) baryonic j-M plane and (b) gas j-M plane. The stability relation (Eqn.~\ref{eqn:toomre} ) is shown as a green dashed line; it can be seen that it does indeed provide a reasonable fit to the data. For the best fit relation shown in Fig.~\ref{angmom_stability} we assume n=1 (since dwarf irregular galaxies typically have rising rotation curves) and fit for $\sigma/Q_C$. As can be seen from Eqn~\ref{eqn:toomre}, if we assume n=0 the derived $\sigma/Q_C$ will change by a factor of $\sqrt{2}$.  The $\sigma$/$Q_{c}$ that we get (for n=1) is $\sim$ 2.8 \kms.  This value can be compared with the canonical values of $\sigma$ and the $Q_{c}$. The observed velocity dispersion in gas disks is $\sim 6 - 10$~\kms  \citep[see e.g.][]{shostak84,tamburro09,ianjamasimanana15}. If we assume that $Q_c$ lies in the range $2-4$ (which is a typical value suggested by observations, see e.g. \cite{zasov17} and references therein), then the obtained best-fit values of $\sigma$/$Q_{c}$ lie well within the range suggested by other observations. Finally, we note that this stability model has also been proposed by \citet{obreschkow16} who present observational support for the prediction of the stability model that the atomic gas fraction ($\fatm$) of galaxies to varies as $\fatm \propto j \sigma/(GM)$ (where $\sigma$ is the gas velocity dispersion). 

In the case of the gas $j-M$ measurements, the best fit $\sigma$/Q$_c$ (for n=1) is $\sim$ 2.2 \kms. This fit was done with the slope fixed to 1, as required by Eqn.~\ref{eqn:toomre}. If we fit also for the slope, the best
fit relationship between the gas specific angular momentum  (\jg) and mass (\mg) is:
\begin{equation}
 \log_{10}(\frac{\jg}{10^3 \kms \kpc}) = (0.80 \pm 0.08)\log_{10}(\frac{\mg}{10^{10}\mathrm M_{\odot}}) + (0.37 \pm 0.12)
\label{eqn:jgas}
\end{equation}

\subsection{Comparison with theoretical expectations}
\label{theory}

One of the clear conclusions of this paper is that dwarf galaxies have a higher baryon specific angular momentum compared to an extrapolation of the trends seen for higher mass bulge-less spirals. Similarly,  \citet{butler17} found that dwarf galaxies also have systematically higher \jb\ as compared to the \jb\ $\propto$ M$_{b}^{2/3}$ relation observed for samples which include all spirals independent of bulge fraction. \citet{badry17} study the angular momentum content of isolated galaxies in the FIRE simulations spanning M$_{\ast}$ = 10$^{6-11}$ M$_{\odot}$. They find that the simulation produces fewer high-angular momentum, low mass galaxies than are represented in observational studies. They suggest that this is due to (1) the standard practice of assuming that the stellar disc has the same rotational velocity as the gas, which leads to an overestimation of angular momentum, and (2) the bias in observational studies towards well behaved, rotating systems. For our sample galaxies, we note that the first issue is unlikely to be a problem since the contribution of stellar specific angular momentum to the total specific angular momentum is very small (< 10 $\%$). We also note that none of the galaxies in the FIRE simulations have as high specific angular momentum as that of the observed galaxies. Thus while it is possible that the relationship that we observe defines the upper envelope of the \jb\ - \mb\ relation, it does appear that even those simulations that include feedback do not produce galaxies with as high a specific angular momentum as found in observations.

The simple tidal torquing model predicts that for the dark matter $j_{H} \propto \lambda M_{H}^{2/3}$. We show in Fig.~\ref{angmom_stability} the corresponding baryonic relation derived for spiral galaxies by \citep{obreschkow14}:
\begin{equation}
\frac{\jb}{10^{3} \ \kpc \ \kms} = 1.96 \lambda f_{j}f_{M}^{-2/3} \Big[ \frac{\mb}{10^{10} \ \msun} \Big]^{2/3} 
\end{equation}
where f$_{M}$ = M$_{b}$/M$_{H}$, is the baryon fraction and f$_{j}$ = j$_{b}$/j$_{H}$ is the specific angular momentum fraction. The derivation by \cite{obreschkow14} assumes f$_{j}=1$ \citep{fall80, stewart13}, spin parameter $\lambda$ $\approx$ 0.04 \citep{maccio08} and f$_{M}$ $\approx$ 0.05 (as typical for large spiral galaxies \citep{mcGaugh10, behroozi13}). As can be seen from Fig.~\ref{angmom_stability}, the relationship predicted by the tidal torquing model is flatter than what we observe for dwarf galaxies.
Of course, as we noted above, this relationship is, strictly speaking, for the total mass, while the observed relationship is for the baryonic components. It seems likely therefore that, as discussed above, baryonic processes that unfold as the galaxy evolves are crucial in determining the final specific angular momentum. In this context, the fact that stability models (as discussed in \S \ref{Stability_model}) predict a j-M relationship quite similar to what is observed is particularly interesting.  In the scenario where the specific angular momentum is regulated by  stability criteria, the larger specific angular momentum of dwarfs may be partly because dwarf galaxy rotation curves are linear over a larger part of the disc. (Recall as discussed above, in the stability model the intercept is 0.29 for a flat rotation curve and 0.44 for a rising one). We note however that our own sample contains a mix of rising as well as flat rotation curves. On the other hand, dwarf galaxies also have significantly thicker gas discs than spirals (\citet{roychowdhury10,roychowdhury13}). Gas column densities that would be unstable by the Toomre criteria (Q=1) for thin discs would be stable in thick ones (see e.g. \citet{wang10}). As such, if the gas column density is being set by the stability criteria, one would expect that this would lower the specific angular momentum for dwarfs as compared to spirals. It hence appears likely that there is something else at play here, apart from disc stability. We explore one such mechanism, viz. preferential loss of low angular momentum gas, in section (\S \ref{ssec:feedback}) below.

\begin{figure}
\centering
\includegraphics[width=1.0\linewidth]{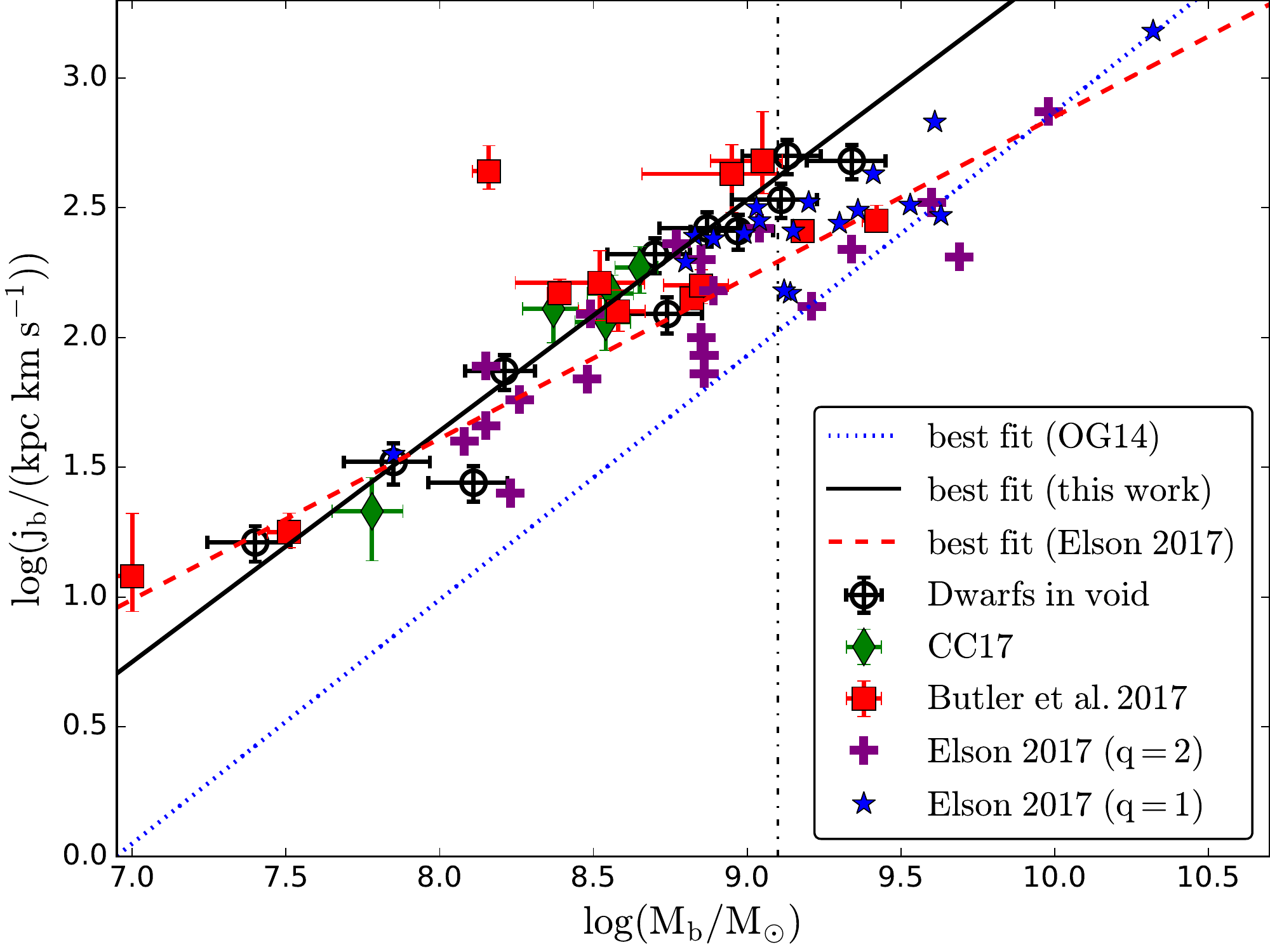}
\caption{  Location of  11 dwarf galaxies residing in Lynx-Cancer void from this work (black open circles), 12 dwarf galaxies from \citet{butler17} (red squares), dwarf galaxies from \citet{chowdhury17} (green diamonds), 17 dwarf galaxies with reliable rotation curves (q=1) from \citet{elson17} (blue stars) and 19 dwarf galaxies with uncertain rotation curves (q=2)from \citet{elson17} (purple plus symbols) in the plot of \jb vs \mb.  The black solid line is the linear regression for dwarf galaxies from this work, the  blue dotted line indicates the $\beta$ = 0 plane of \jb - \mb relation obtained for bulgeless spiral galaxies by \citet{obreschkow14,chowdhury17} and the red dashed line is the best fit relation obtained by \citet{elson17}.}
\label{fig:elson}
\end{figure}%

\begin{figure}
\centering
\includegraphics[width=1.0\linewidth]{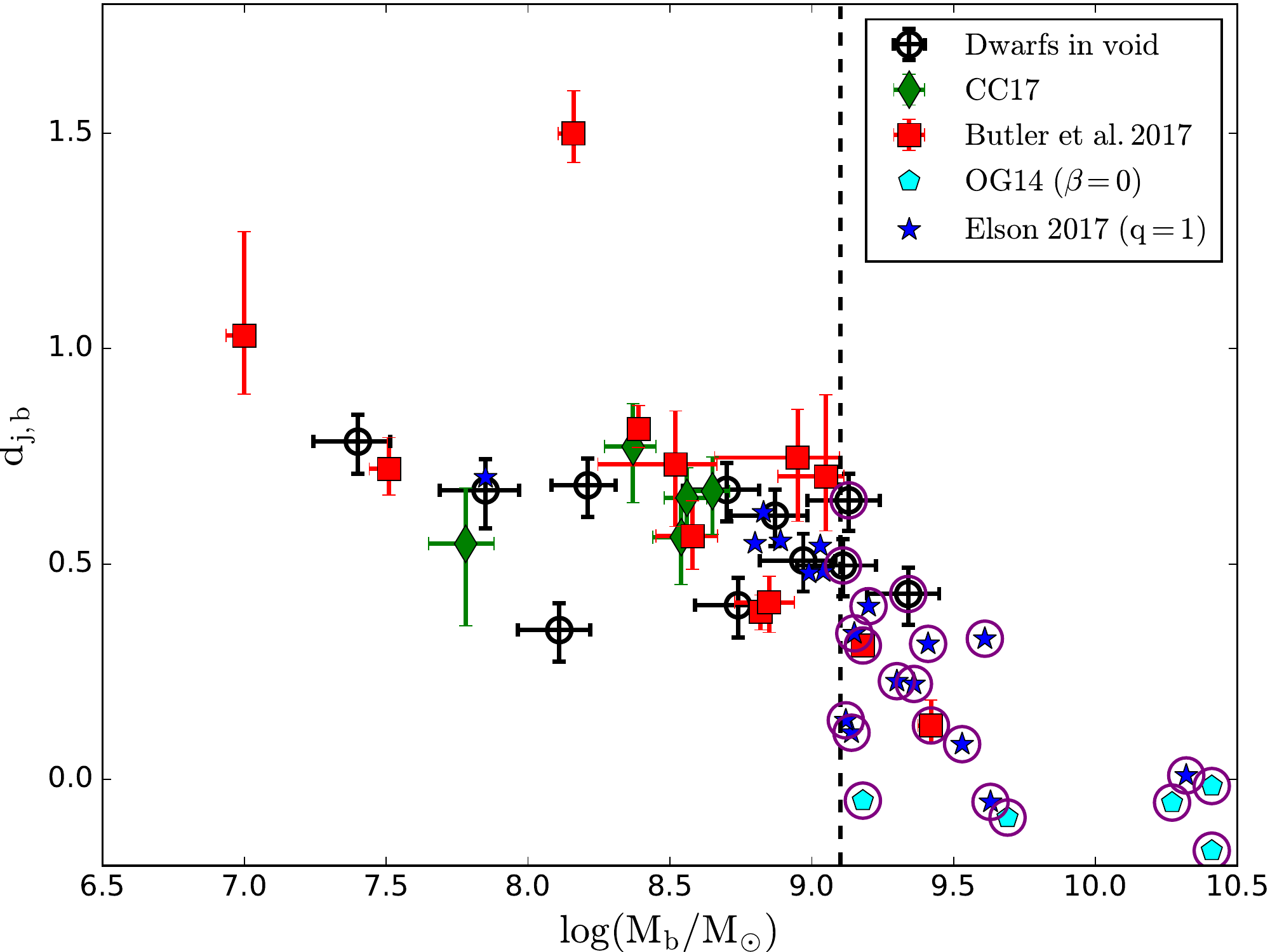}
\includegraphics[width=1.0\linewidth]{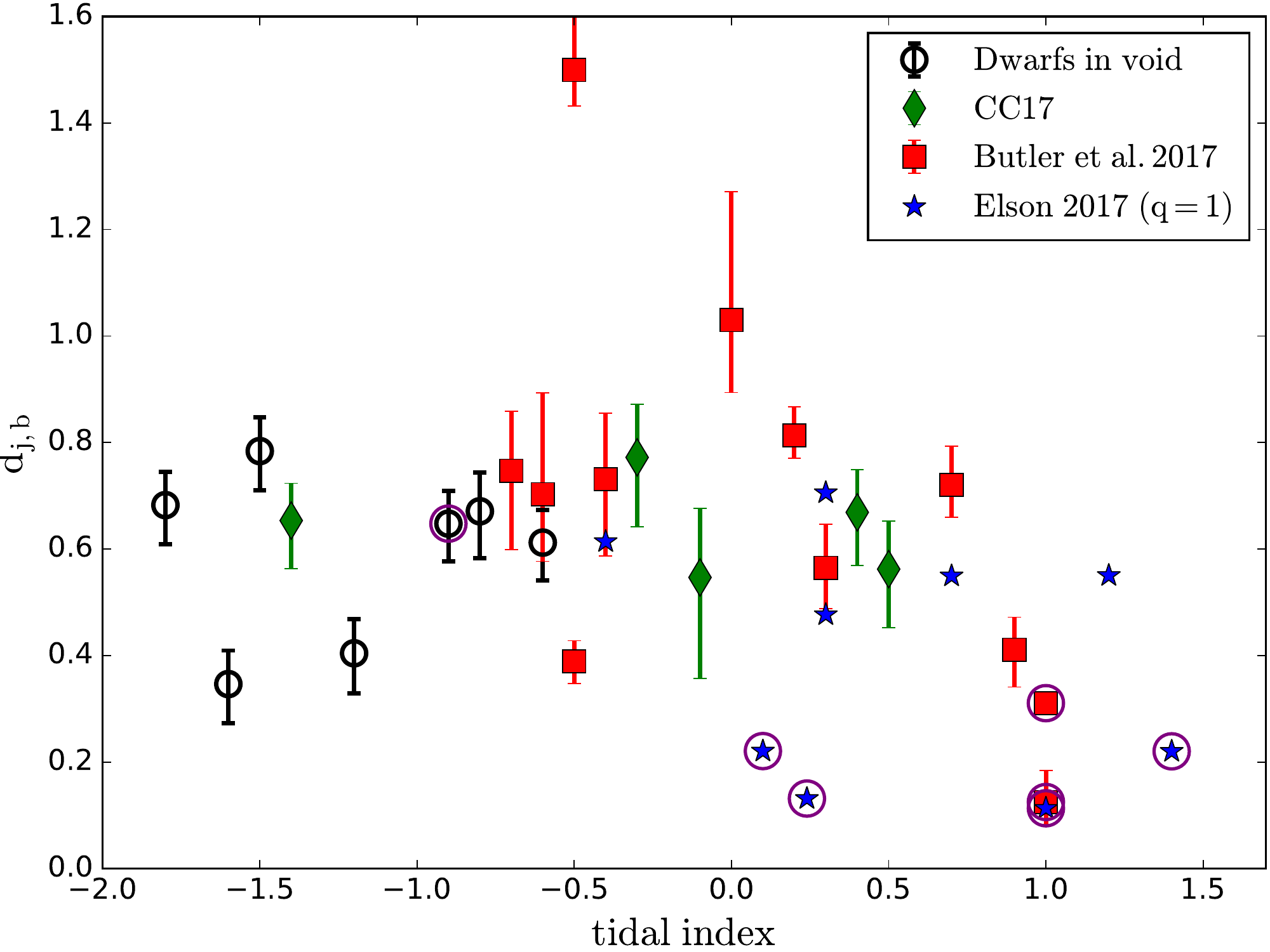}
\caption{ (a) d$_\mathrm{j,b}$  versus baryon mass and (b) d$_\mathrm{j,b}$  versus tidal index.  11 dwarf galaxies residing in Lynx-Cancer void from this work (black open circles), and the dwarf galaxies from the literature (symbols are the same as in previous figures). 5 massive spiral galaxies with bulge fraction, $\beta$ < 0.05 from \citet{obreschkow14} are shown by cyan pentagons. The galaxies with masses higher than 10$^{9.1}$ M$_{\odot}$ are indicated by a circle around their respective symbols.}
\label{fig:mbtidal}
\end{figure}%

\subsection{Comparison with other samples}

In this section, we compare the \jb\ - \mb\ relation from this work to the relation obtained for dwarf galaxies in earlier studies. As discussed above, our results are in good agreement with those of \cite{butler17} and \cite{chowdhury17}.  Recently \cite{elson17} used a sample of 37 galaxies from the Westerbork H{\sc i} survey of Spiral and Irregular Galaxies project (WHISP; \citep{swaters02}) with rotation curve data from  \citet{swaters09} to derive the \jb\ - \mb\  relation. \citet{swaters09}  used data with 30$^{"}$ resolution to derive the rotation curves and they divided their derived rotation curves into four categories. A reliable rotation curve was given a quality index q=1, uncertain and highly uncertain rotation curves are indicated by q=2 and q=3 respectively. A case for which no rotation curve could be derived is indicated by q=4. \citet{elson17} use both reliable rotation curves (q=1) and uncertain rotation curves (q=2) for their study of \jb\ - \mb\ relation for low mass galaxies.  Fig. \ref{fig:elson} shows the location of dwarf galaxies from this work and dwarf galaxies from the literature in the plot of \jb versus \mb. The black solid line is the linear regression for dwarf galaxies from this work, the blue dotted line indicates the $\beta$ = 0 plane of \jb - \mb relation obtained for the massive spiral galaxies by \citet{obreschkow14,chowdhury17} and the red dashed line is the best fit relation obtained by \citet{elson17}.  \citet{elson17} obtained a slope, $\alpha$ $\sim$ 0.62, which differs from the steeper relation obtained in this work ( $\alpha$ $\sim$ 0.89). However, if we consider only the galaxies with reliable rotation curves (q=1), and with masses lower than 10$^{9.1}$ M$_{\odot}$ (see Sec.~\ref{djb} below for a justification of this cut-off), then the galaxies  seem to be consistent with the relation obtained by us. The galaxies with uncertain rotation curves (q=2) have a large scatter and deviate from the distribution of remaining dwarf galaxies. As discussed below, combining our sample with other samples, we find a break in the \jb - \mb relation at a mass of 10$^{9.1}$ M$_{\odot}$.  Most of our sample galaxies are below this mass limit. The best fit slope we find using \citet{elson17} sample, after excluding galaxies with uncertain rotation curves as well as galaxies with masses > 10$^{9.1}$ M$_{\odot}$ is 0.86 $\pm$ 0.04. This agrees within error bars with the value that we find for our sample.

\subsection{Dependence of the j$_{b}-M_{b}$ relation on other galaxy parameters}
\label{djb}

In order to understand if there is any correlation between the deviation of the specific angular momentum of dwarf galaxies from the \jb\ - \mb\ relation of bulgeless spiral galaxies with other parameters such as baryonic mass and environment, we calculate the difference in \jb\ and j$_\mathrm{b,exp}.$ where, \jb\ is the observed specific angular momentum for a dwarf galaxy and j$_\mathrm{b,exp}$ is the expected specific angular momentum for a given mass and zero bulge fraction from the relation derived for spiral galaxies obtained by \citet{obreschkow14,chowdhury17}. Let d$_\mathrm{j,b}$ be defined as log(\jb)-log(j$_\mathrm{b,exp}$). 

Fig. \ref{fig:mbtidal} (a) shows the plot of  d$_\mathrm{j,b}$ versus baryon mass. We include the dwarf galaxies from this work and dwarf galaxies from the literature.  However, we include only the galaxies with reliable rotation curves (q=1) from \citet{elson17}.  A break appears around \mb  $\sim$ 10$^{9.1}$  M$_{\odot}$. Dwarf galaxies with masses lower than 10$^{9.1}$  M$_{\odot}$ have significantly higher baryonic specific angular momentum than expected from the relation for bulgeless spiral galaxies. As the mass of the galaxy increases beyond 10$^{9.1}$  M$_{\odot}$, the baryonic specific angular momentum decreases and they tend to follow the relation obtained for the bulgeless spirals. It is interesting to note that the mass at which the break is seen is similar to the limiting mass M$_{\ast}$ $\sim$ 2 $\times$ 10$^{9}$ M$_{\odot}$, below which low-mass galaxies start to be systematically thicker \citep{janssen10}.

We also check if there is any correlation between the scatter in \jb\ - \mb\ relation with the environment. We use the tidal index (TI) as a parameter to describe the environment of a galaxy. Tidal index quantifies the local density environment and is determined by the distance and the mass of significant neighbour.  We use TI5 as a parameter which is determined by the five most important neighbours. TI5 is a more robust indicator of galaxy environment as compared to TI1, which is determined by the galaxy's immediate significant neighbour. Negative tidal indices indicate that the galaxy is in an isolated environment while large positive indices indicate that the galaxy is in a dense environment. The TI values were primarily taken from the Local Volume (LV) catalogue \citep{karachentsev14}; for some of the galaxies with significant changes in the estimated distance, they were recomputed and corrected using the updated information in the extragalactic database, HyperLeda. Fig.~\ref{fig:mbtidal}~(b) shows the plot of d$_\mathrm{j,b}$ versus tidal index. There is some indication that galaxies with higher tidal indices also have lower specific angular momentum, however, the trend is not as clear as it is in the case of the mass (Fig.\ref{fig:mbtidal}~(a)). Further, for our sample, the galaxies with larger tidal index, also tend to be the ones with higher baryonic mass. We have indicated all the galaxies with masses higher than 10$^{9.1}$ M$_{\odot}$ by a circle around their respective symbols. In Fig.~\ref{fig:mbtidal}~(b), we do not see any trend for specific angular momentum to decrease with higher tidal index if we ignore the high-mass galaxies (>10$^{9.1}$ M$_{\odot}$).

\subsection{Preferential loss of low angular momentum gas via stellar feedback}
\label{ssec:feedback}

One possibility discussed by \cite{chowdhury17} is that the specific angular momentum in the dwarfs has been increased by preferential loss of low angular momentum gas. This is a natural consequence of stellar feedback, in situations where the star formation is concentrated near the galaxy center (i.e. a region of low specific angular momentum) and the mass loss happens preferentially along the minor axis of the galaxy \citep[see e.g.][]{brook11}. Such situations arise naturally in dwarf galaxies, and result in stellar feedback processes preferentially removing low angular momentum gas from the central parts of dwarfs, thus increasing the specific angular momentum of the remaining gas. Mechanical energy input from stellar feedback would also increase the thickness of the disk. While this qualitatively looks like an attractive explanation, it would be interesting to check if we can confirm that the amount of star formation in these galaxies is sufficient to produce the observed increase in the specific angular momentum. This is a particular concern for the extremely gas rich galaxies in our sample, since they have a small fractional stellar mass, and it is unclear that the star formation would be sufficient to change the specific angular momentum by the observed factor.

To estimate the expected increase in the specific angular momentum because of preferential loss of low specific angular momentum gas, we firstly assume that the \jb\ - \mb\ curves for the bulgeless spirals and dwarfs are approximately parallel (since the measured slopes agree with error bars). If we call the factor by which the dwarfs specific momentum is larger than that of the spirals as $\beta$, then the offset between the lines in Fig.~\ref{angmom} corresponds to
$\log_{10}(\beta) = 0.53 \pm 0.09$, which  corresponds to $\beta = 3.4 \pm 0.21$. For the purpose of making an approximate estimate of the required mass loss, we assume, that to zeroth order the disks are marginally stable, so that their mass and angular momentum are given by Eqn.~\ref{eqn:mgcrit} and ~\ref{eqn:Jcrit} respectively. If we assume that outflows remove all of the material within a fractional radius $\alpha$, (i.e. all of the material between the galaxy centre and $\alpha R_0$ is lost), then from these same equations one can work out that the mass will decrease by the factor $(1-\alpha)$ and the offset in \jb\ - \mb\ plane will increase by the factor $\log_{10}((1+\alpha)/(1-\alpha))$. This allows us to solve for $\alpha$,
yielding $\alpha = 0.55 \pm 0.07$, or that about half of the original baryonic mass needs to be lost. From the simulations including stellar feedback \citep[e.g.][]{brook11}, the mass loading factor (i.e. the ratio of the mass loss rate
to the star formation rate) is $\sim 2.3$. This is also the ratio between the total mass lost and the total stellar mass. From the data in Table~\ref{table2} the average M$_{*}$/M$_{b}$ is $\sim 0.08$, which for a mass loading factor of 2.3 corresponds to a total fractional mass loss of $0.16$,
significantly smaller than the value of $\alpha = 0.55 \pm 0.07$ required to explain the offset in the \jb\ - \mb\ plane. We note that the assumption that all of
the material internal to $\alpha R_0$ is lost is an overestimate, and hence
our calculation of the increase in the specific angular momentum is also
an overestimate. Thus the conclusion that the observed star formation is insufficient to increase the specific angular momentum by the observed amount appears robust. On the other hand, the mass loading factor includes only mass that escapes from the galaxy altogether. In principle, it is sufficient that the mass is converted into a form (e.g. ionized gas) that drops out from our baryon budget.  While this is a possible solution, there is no clear observational evidence that the ionized medium contains a significant fraction of the baryonic mass in dwarf galaxies.

Given that there appear to be problems with increasing the specific angular momentum entirely by preferential loss of low angular momentum gas, it is worth considering what other possible mechanisms could be relevant. One such, suggested earlier by \citet{chowdhury17} in the context of the high specific angular momentum of dwarfs, is the cold accretion of high angular momentum gas (see also e.g. \cite{danovich15}). This mechanism, if operative, would of course result in increasing the specific angular momentum of dwarf discs. One could speculate that it would result in an increased disk thickness because of increased turbulence caused by the accretion, but it is unclear as to why cold accretion should be more relevant for dwarfs as compared to bulgeless spirals.

\section{Summary and Conclusions}
\label{summary}

In summary, we have measured the specific angular momentum of 11 void dwarf galaxies and find them to be larger than would be expected from an extrapolation of the trends seen for bulgeless spirals. We compare the specific angular momentum of dwarf galaxies that lie in voids with that of dwarfs that lie outside voids and find that they have similar specific angular momentum(\S \ref{j-voids}). As such, our data does not show any correlation between the specific angular momentum and the large scale environment. We hence derive the relation between the baryonic specific angular momentum and mass for dwarf galaxies by combining our data with that for other dwarfs for which data is available in the literature(\S \ref{M-j}). We find that the dwarf galaxies with masses lower than $\sim 10^{9.1}$ M$_\odot$ have significantly higher baryonic specific angular momentum than expected from the relation for bulgeless spiral galaxies (\S \ref{djb}).  
Interestingly this threshold is very similar to the mass threshold below which the galaxy discs start to become systematically thicker \citep{janssen10}. This finding gives qualitative support to earlier suggestions that both of these processes may arise because of a common physical mechanism, such as feedback from star formation. Stellar feedback processes preferentially remove low angular momentum gas from the central parts of dwarfs (thus increasing the specific angular momentum of the remaining gas)  and the input mechanical energy also leads to an increase in the velocity dispersion and to a thickening of the discs. In detail, however, we find that the observed amount of star formation in our sample galaxies is insufficient to produce the observed increase in the specific angular momentum (\S \ref{theory}). We suggest that other mechanisms, such as for example, cold mode accretion of high specific angular momentum gas, may also be playing a role. In any case, our findings underline the importance of the specific angular momentum as a tool to constrain galaxy formation scenarios.

\section*{Acknowledgements}

SAP appreciate the support of this work through
RSCF grant No. 14-12-00965. The authors thank
Y. Perepelitsyna for the help with photometry of
two void galaxies outside the SDSS footprint. We acknowledge the usage of the HyperLeda database (http://leda.univ-lyon1.fr). This paper is based in part
on observations taken with the GMRT. We thank the staff of the
GMRT who made these observations possible. The GMRT is run
by the National Centre for Radio Astrophysics of the Tata Institute
of Fundamental Research.




\bibliographystyle{mnras}
\bibliography{angmom} 





\appendix

\section{Rotation curves}
\begin{figure}
\centering
\includegraphics[width=0.9\linewidth]{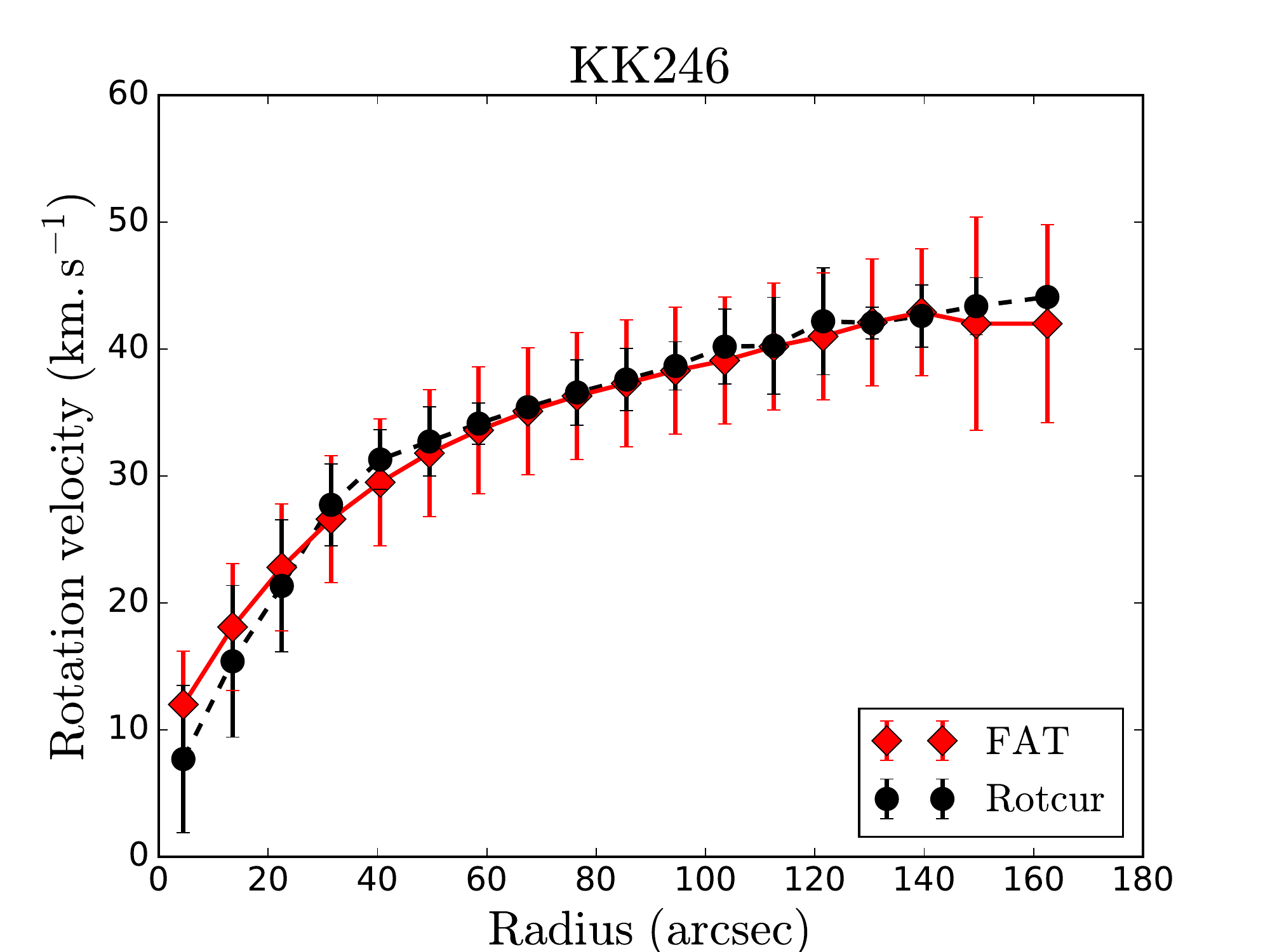}
\includegraphics[width=0.9\linewidth]{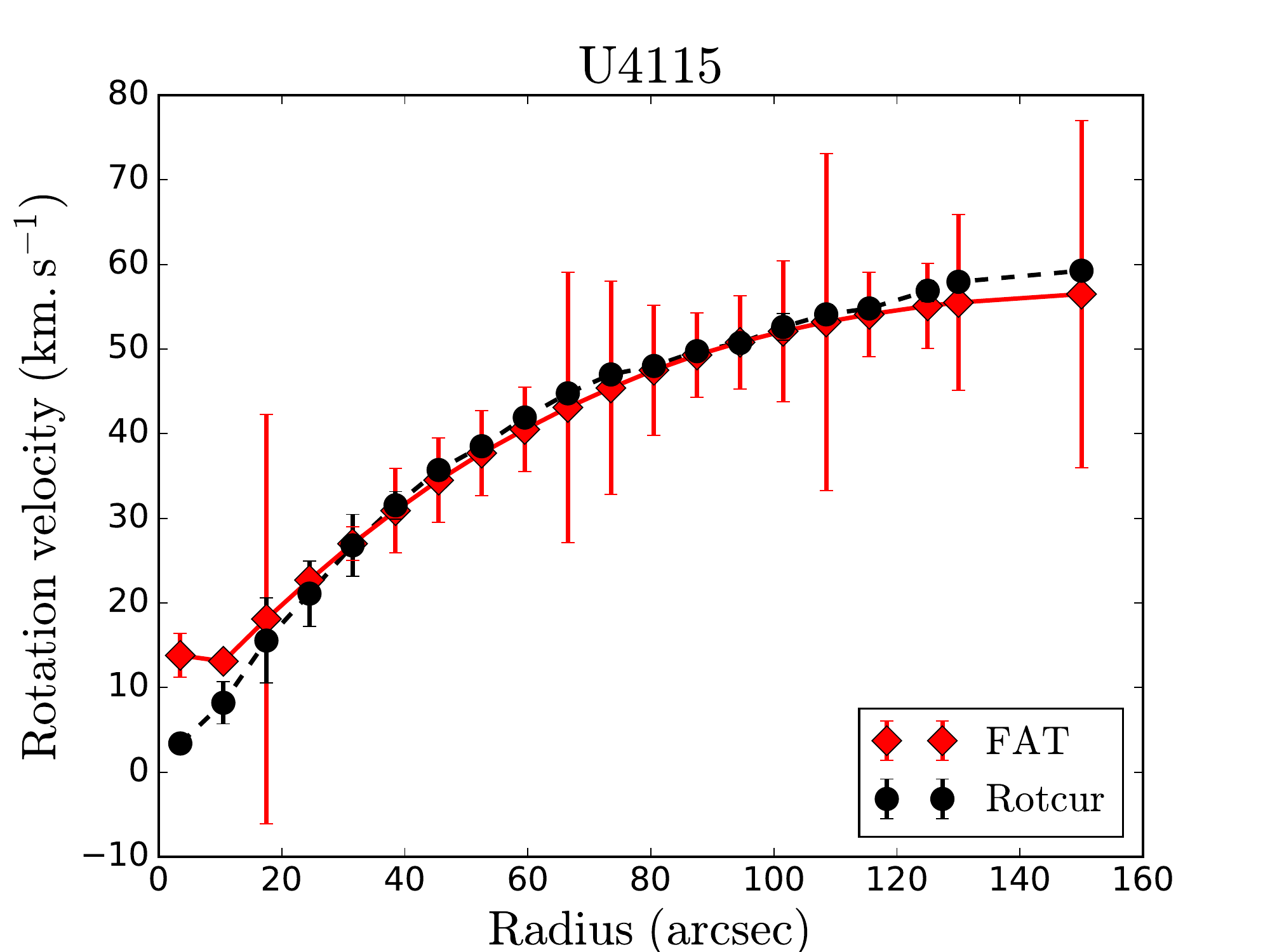}
\includegraphics[width=0.9\linewidth]{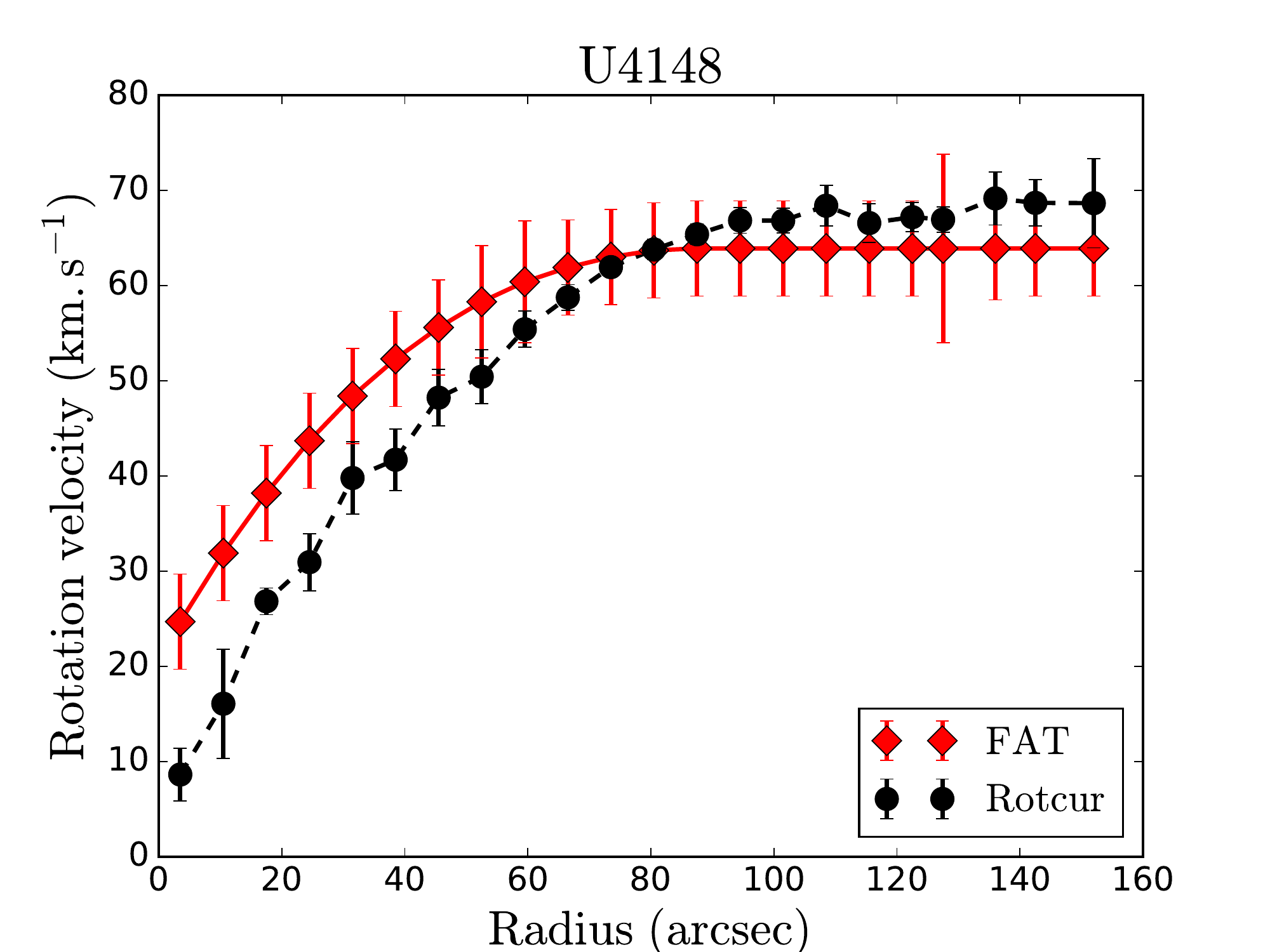}
\includegraphics[width=0.9\linewidth]{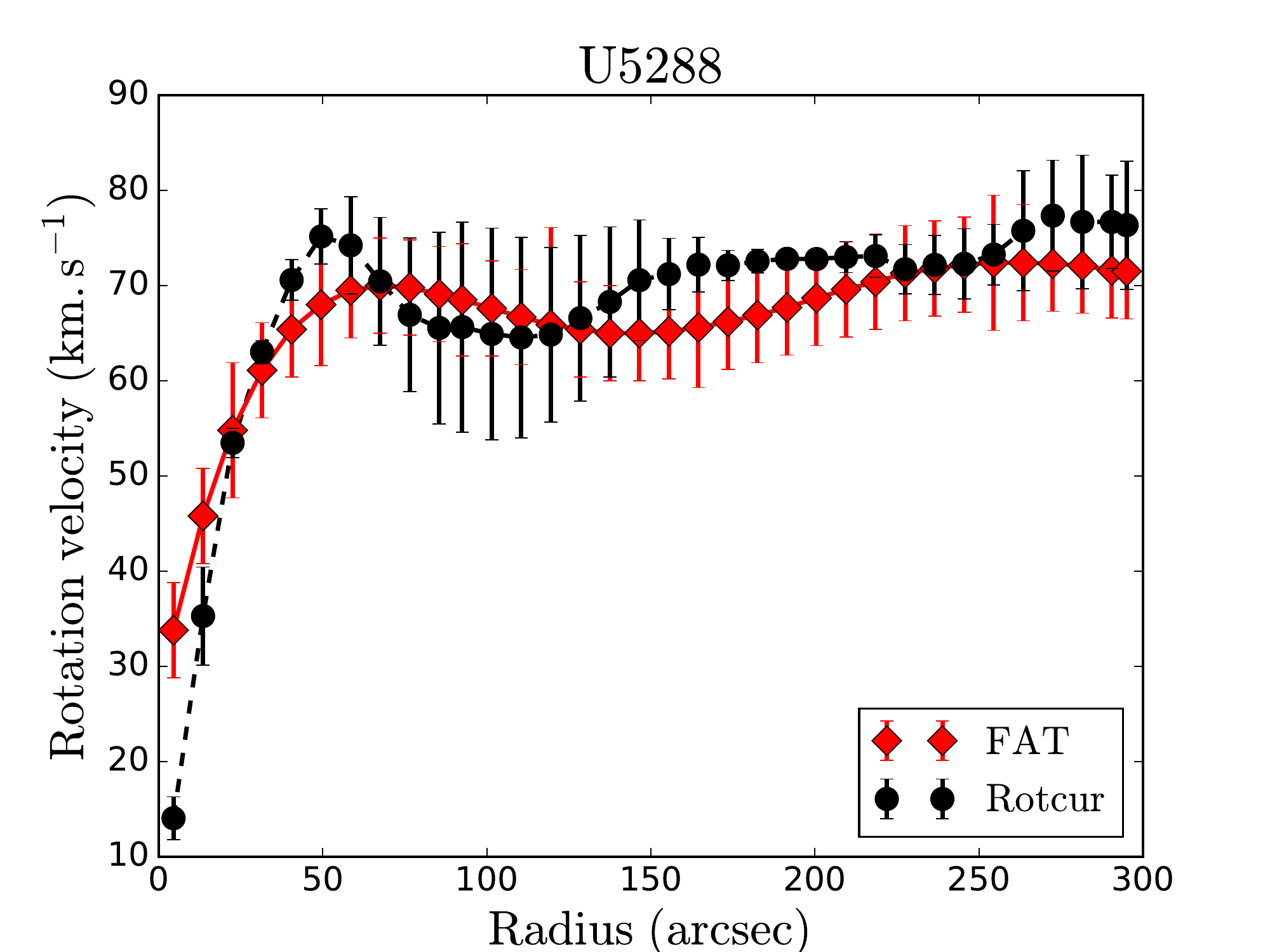}
\caption{Rotation curves of galaxies, where rotation curves from FAT pipeline match with that of the ROTCUR software.}
\label{rotfat1}
\end{figure}%
\begin{figure}
\centering
\includegraphics[width=0.9\linewidth]{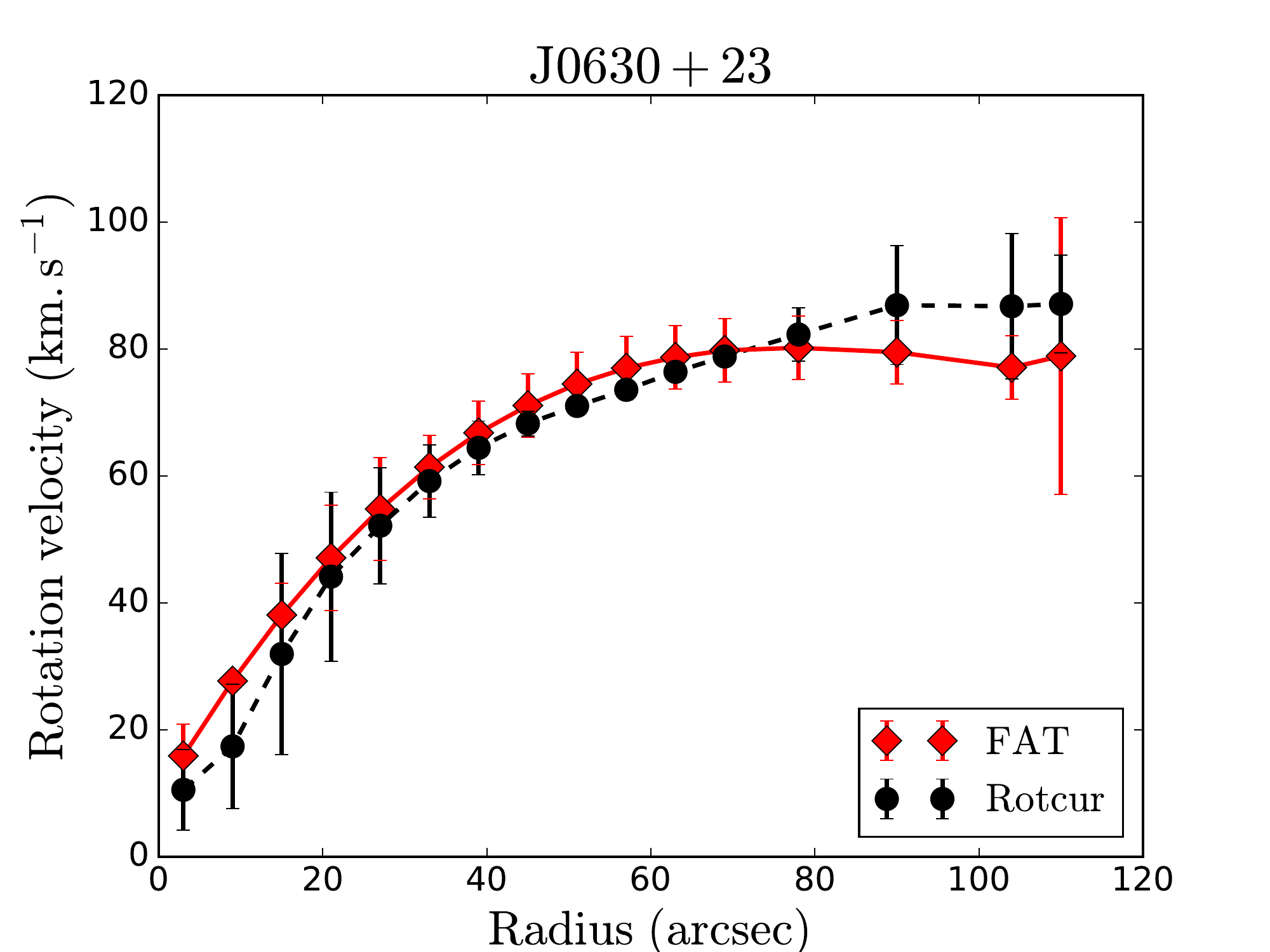}
\includegraphics[width=0.9\linewidth]{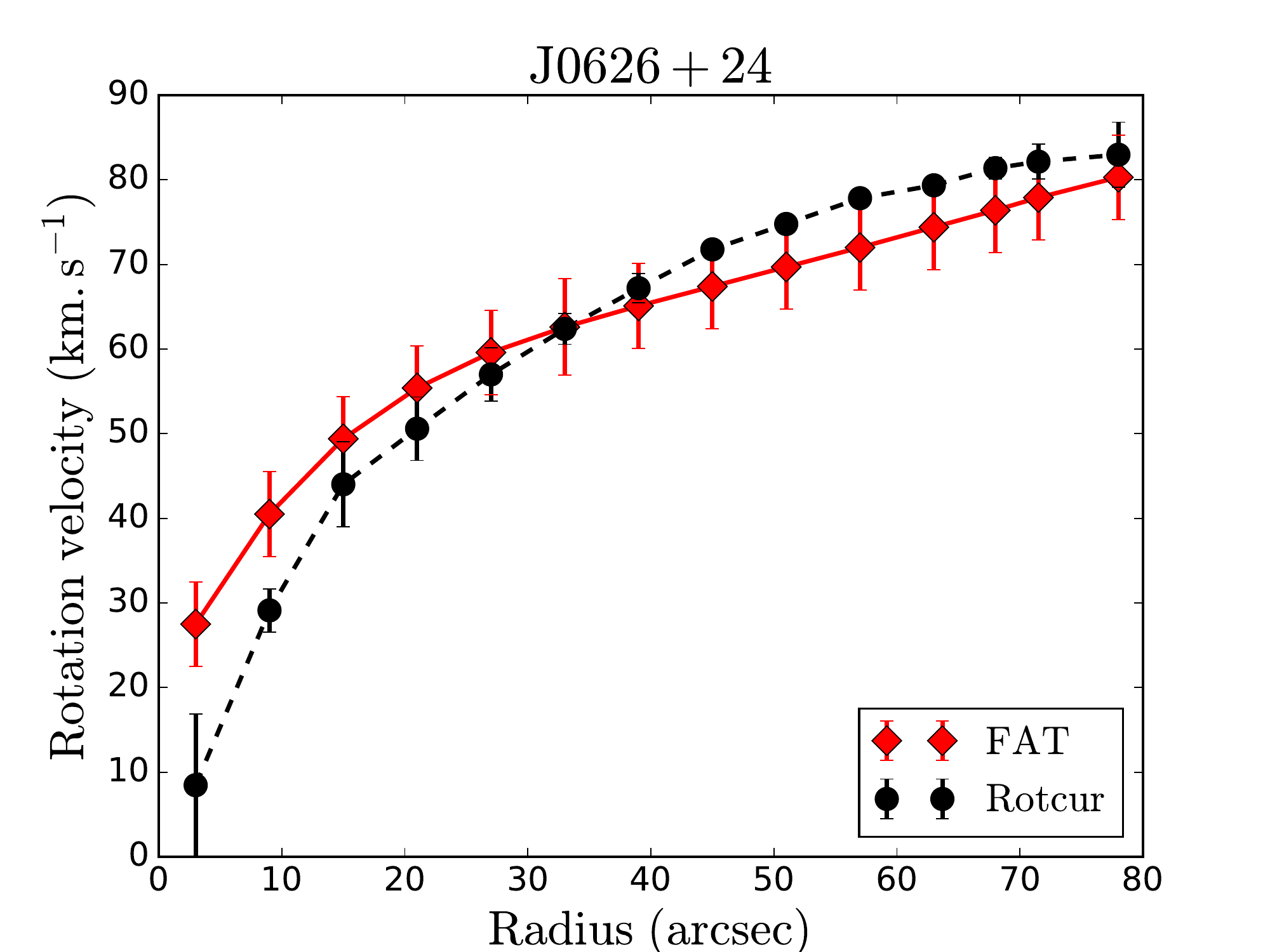}
\includegraphics[width=0.9\linewidth]{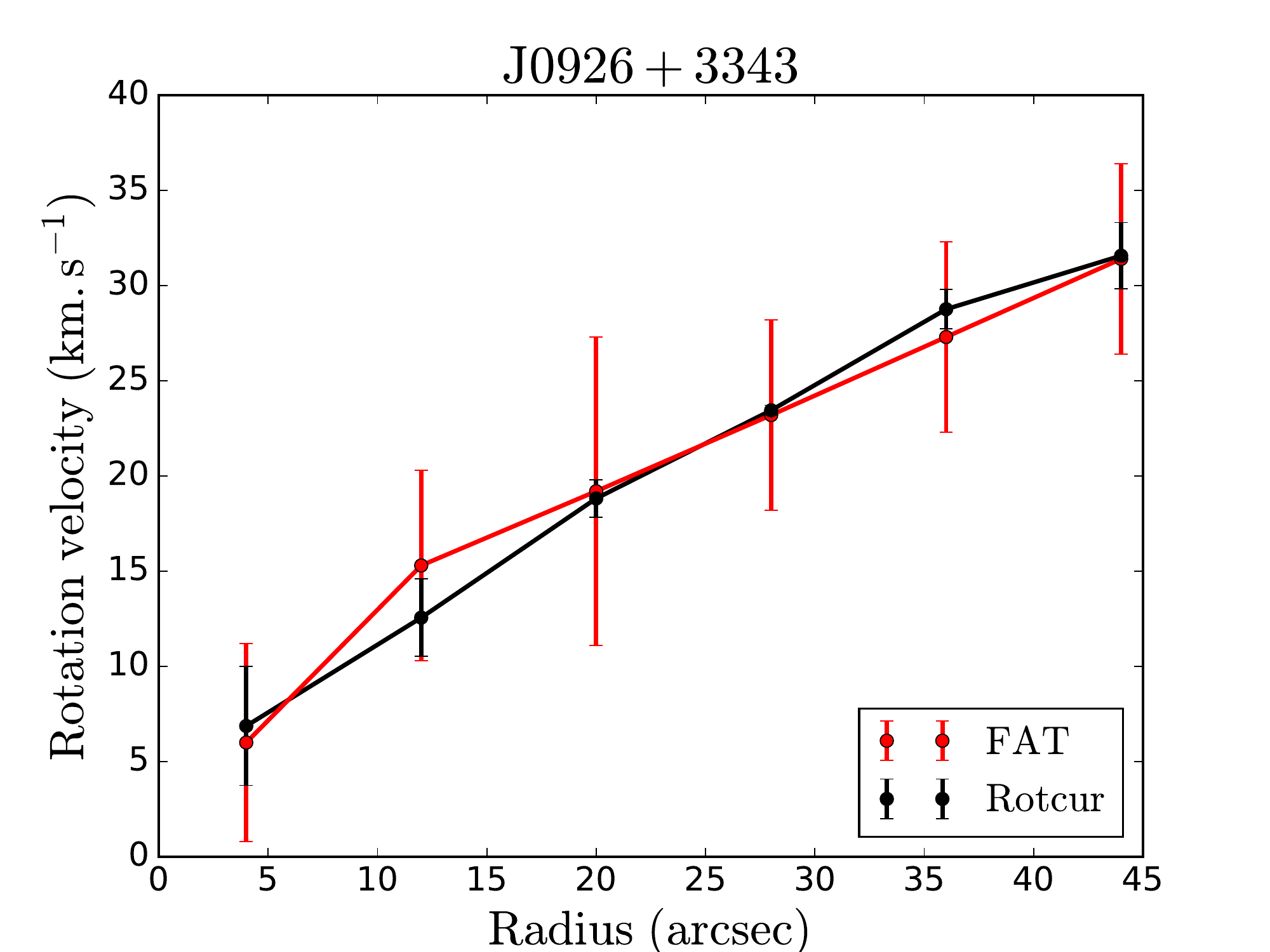}
\includegraphics[width=0.9\linewidth]{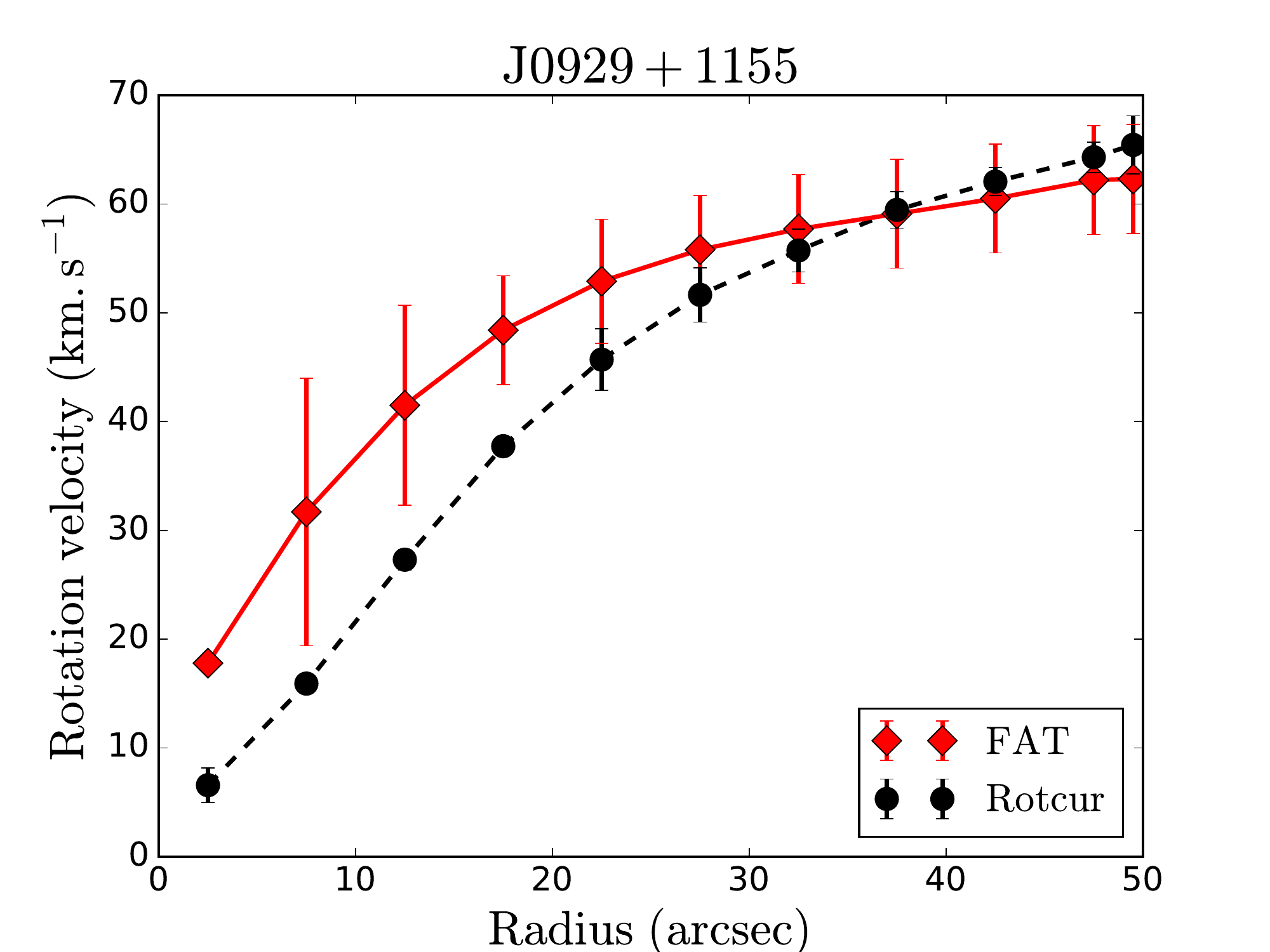}
\caption{Rotation curves of galaxies, where rotation curves from FAT pipeline match with that of the ROTCUR software.}
\label{rotfat2}
\end{figure}
\begin{figure}
\centering
\includegraphics[width=0.9\linewidth]{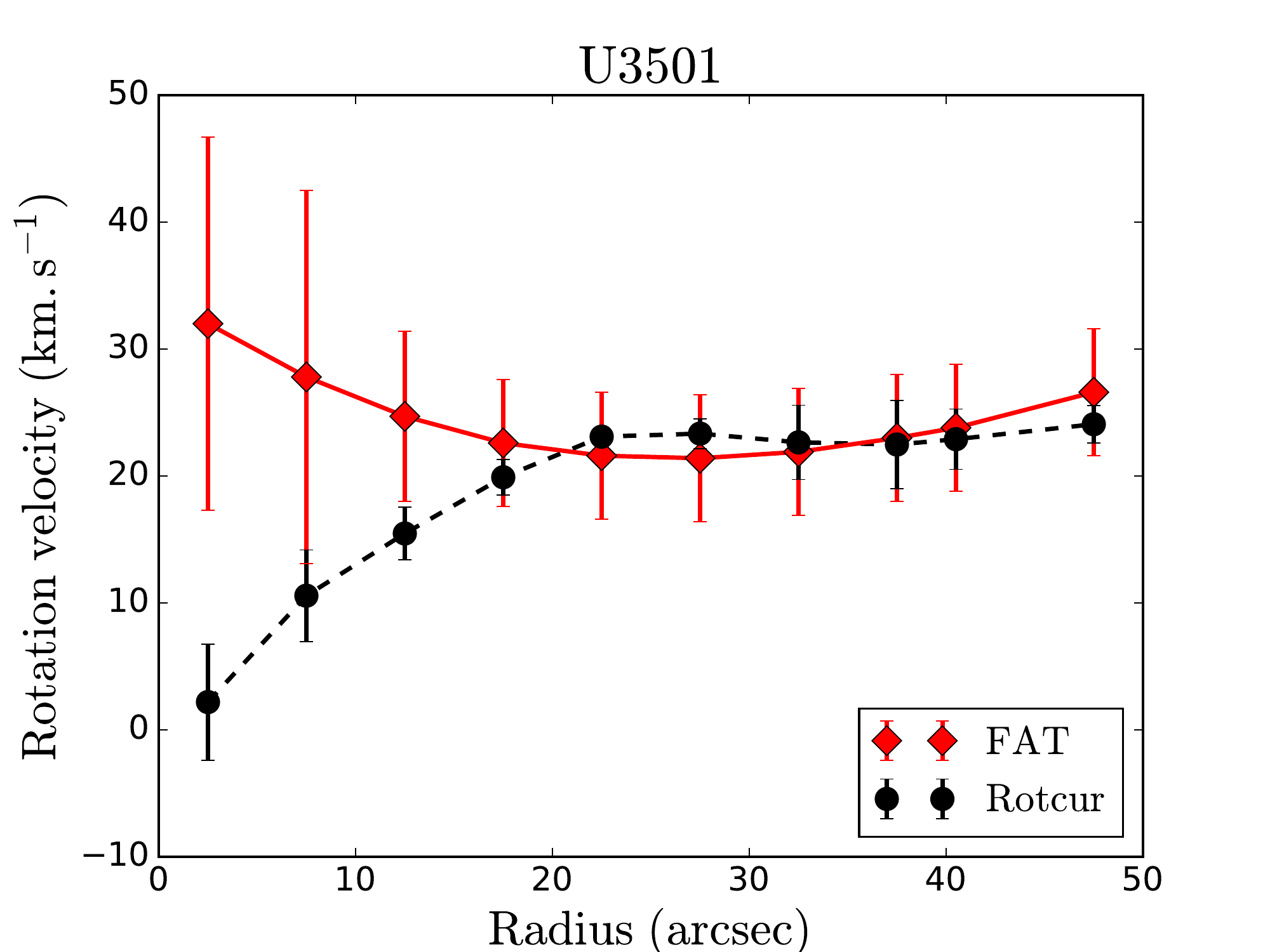}
\includegraphics[width=0.9\linewidth]{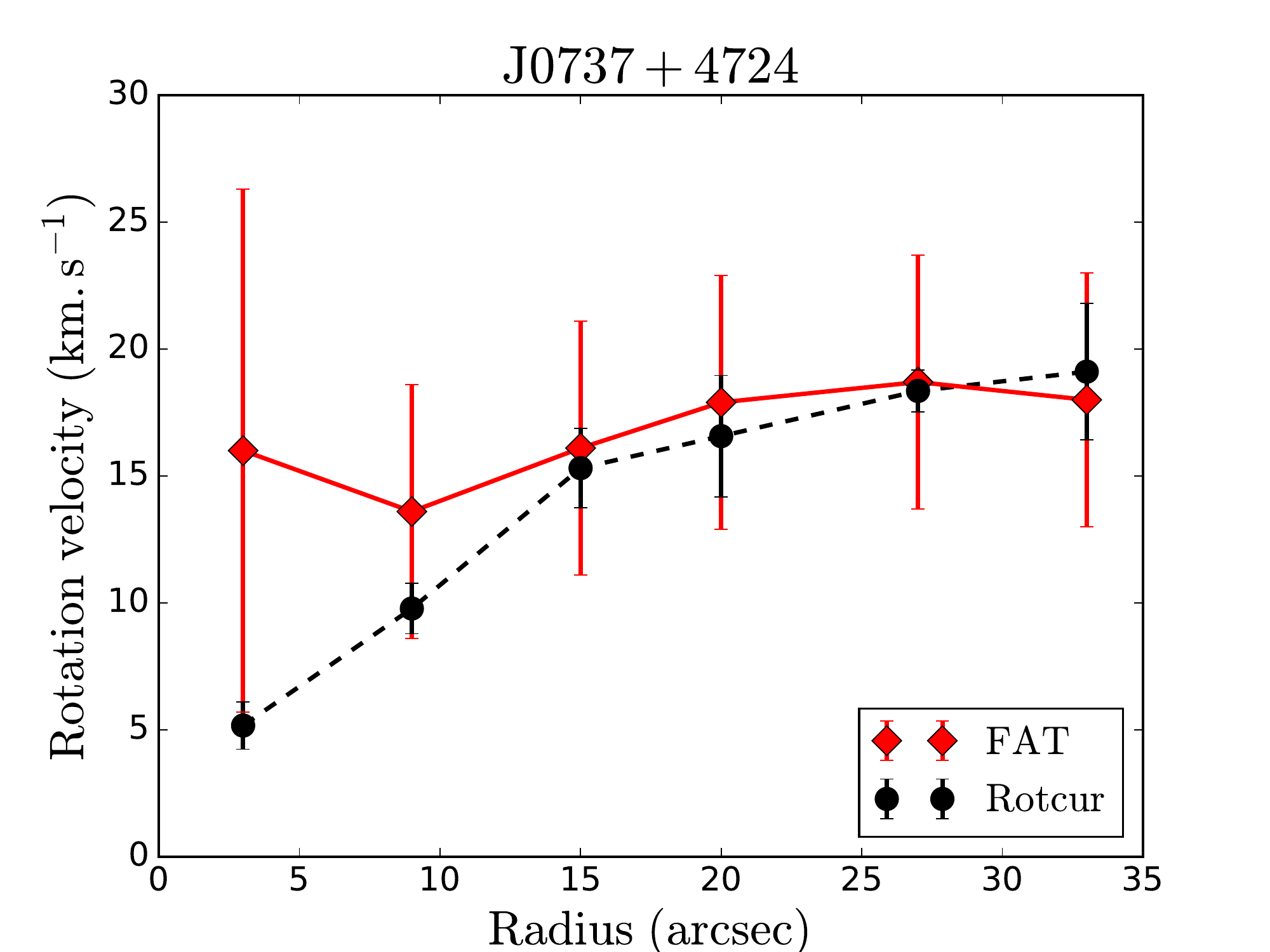}
\includegraphics[width=0.9\linewidth]{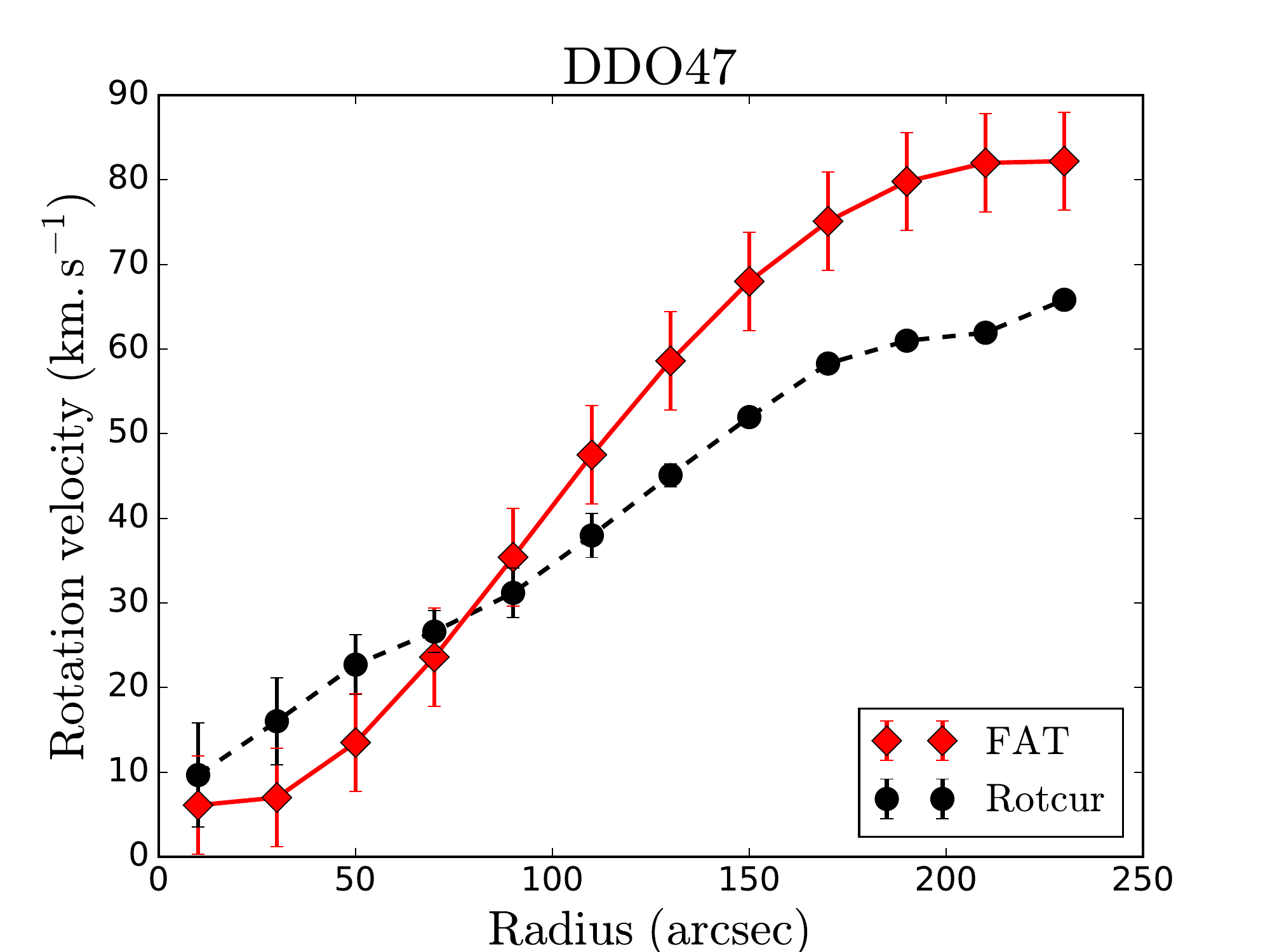}
\caption{Rotation curves of 3 galaxies, where reliable rotation curves could not be obtained using the FAT pipeline.}
\label{rotfat3}
\end{figure}

%
Here, we present the rotation curves derived using the FAT, which fits the 3-D tilted ring model to the data cubes and using the ROTCUR, which fits the 2-D tilted ring model to the velocity fields (see \S \ref{rotcur} for more details). Fig. \ref{rotfat1}, \ref{rotfat2} and \ref{rotfat3} show the comparison of rotation curves of 11 dwarf galaxies obtained using the FAT pipeline and the ROTCUR. The rotation curve obtained using ROTCUR is shown with black circles and a black dotted line and the rotation curve  from the FAT pipeline is shown with red diamonds and a red solid line. Fig. \ref{rotfat1} and Fig. \ref{rotfat2} show the rotation curves of 8 galaxies, where the rotation curves derived with the FAT mostly match with that of the rotation curves derived with the 'ROTCUR'. Fig. \ref{rotfat3} shows the rotation curves of 3 galaxies (J0737+4724, UGC 3501 and DDO47), where reliable rotation curves could not be obtained using the FAT pipeline.

\bsp	
\label{lastpage}
\end{document}